\documentclass[12pt]{amsart}

\usepackage{setspace,graphicx,srcltx,enumitem,harvard,bm,xcolor}

\newtheorem{assumption}{Assumption}[section]
\newtheorem{definition}{Definition}[section]

\title[]{Recent Advances in Causal Analysis of the Stochastic Frontier Model}

\date{\today}

\author{Samuele Centorrino \and Christopher F.~Parmeter}

\thanks{Christopher F. Parmeter, University of Miami, Miami, FL, 33146, e-mail: cparmeter@bus.miami.edu. Samuele Centorrino, Bates White LLC and Toulouse School of Economics, email: scentorrino@proton.me. All errors in this paper are due to a violation of parallel trends. We thank Kien Tran, Oleg Badunenko and Giovanna D'Inverno for feedback on earlier versions of this work. All errors are ours alone.}


\begin{document}

\begin{singlespacing}
 \begin{abstract}
Causal inference methods (instrumental variables, difference-in-differences, regression discontinuity, etc.) are primary tools used across many social science milieus. One area where their application has lagged however, is in the study of productivity and efficiency. A main reason for this is that the nature of the stochastic frontier model does not immediately lend itself to a causal framework when interest hinges on an error component of the model. This paper reviews the nascent literature on attempts to merge the stochastic frontier literature with causal inference methods. We discuss modeling approaches and empirical issues that are likely to be relevant for applied researchers in this area. This review shows how this model can be easily put within the confines of causal analysis, reviews existing work that has already made inroads in this area, addresses challenges that have yet to be met and discusses core findings. 
 \end{abstract}
\end{singlespacing}


\maketitle

\onehalfspacing


\section{Introduction}

Stochastic frontier methods (SFA) continue to be one of the workhorse tools for efficiency evaluation across a range of industries. The approach is simple to understand and easily portable across applications. The appeal of decomposing deviations from a frontier into those that are noise (and entirely outside of control of the firm) and those that capture inefficiency offers many avenues to study how performance can be improved. 

What has been of interest lately has been the inevitable consequence of endogeneity. Approaches to dealing with endogeneity in stochastic frontier models have started to reach maturity. Beginning with \citeasnoun{KUTLU:2010} and subsequently studied in \citeasnoun{TRAN_TSIONAS:2013}, \citeasnoun{AMSLER_PROKHOROV_SCHMIDT:2016}, \citeasnoun{AMSLER_PROKHOROV_SCHMIDT:2017}, \citeasnoun{KARAKAPLAN_KUTLU:2017}, \citeasnoun{KUTLU_ETAL:2020}, \citeasnoun{CENTORRINO_PEREZ:2019}, and \citeasnoun{CENTORRINO_PEREZ_URETA_WALL:2024}, an array of approaches have been put forth to disentangle causal effects of inputs from those that are due to structural issues tied to selection and other effects from noise. However, a central limitation of this class of methods is their reliance on the availability of valid instruments, which can be akin to locating the end of a rainbow. While one may believe its position can be identified, the destination is ultimately illusory, always remaining just out of reach. Likewise, \textit{prima facie}, many instruments appear to satisfy exclusion restrictions and generate exogenous variation in the endogenous variable, yet these claims are typically argued rather than formally tested.

And so, recently, researchers have turned to causal inference methods, primarily regression discontinuity designs (RDDs) and difference-in-differences (DiD) to parse out structural effects in stochastic frontier settings. These methods differ from traditional instrumental variable frameworks because they rely on more clearly defined policy changes that lead to exogenous variation in key variables that drive production/efficiency. At the same time, it appears that recent attempts to merge DiD/RDD with SFA have not fully embraced the structure of the model. For example, in \citeasnoun{DINVERNO_ETAL:2023} they seek to exploit a new balanced budget rule, known as the ``policy and management cycle'' which started in 2014 by the local government in Flanders, Belgium. They first use the endogenous stochastic frontier model of \citeasnoun{KARAKAPLAN_KUTLU:2017} and \citeasnoun{KUTLU_ETAL:2020} to recover both the local government technology and efficiency. They then take this estimate of efficiency and run a separate DiD regression model. This framework falls within the well-known two-step approach of efficiency analysis as originally documented in \citeasnoun{WANG_SCHMIDT:2002} and further discussed in \citeasnoun{PARMETER_KUMBHAKAR:2014}. A potential concern is that the DiD structure is ignored when the efficiency scores are first constructed and so when the DiD regression is run in a second stage two problematic issues may arise. First, the efficiency scores are compressed more than they otherwise would be because the average estimator is a shrinkage estimator, compressing towards the overall mean; and, second, the failure to account for the DiD structure at the onset leads to an omitted variable type phenomena, so that even if the endogeneity corrected stochastic frontier is deployed the DiD structure is not present. There is also a third issue which is that the DiD structure is assumed to only impact inefficiency but not technology. 

We also mention that a recent paper by \citeasnoun{TO-THEA_TUAN:2019} proposed an augmented approach where they estimated the stochastic frontier model for each of the four pieces of the DiD (pre-treatment, pre-control, post-treatment, post-control), but their setup may also fail because they do not keep the pre/post control groups the same, which is what one would anticipate when using the DiD since only the shift of the treatment group should arise. \citeasnoun{KARANKI_LIM:2024} used a slightly less rigorous approach and included the DiD structure into the frontier itself, but left the error term untouched. In this case the argument is that the policy change does not impact inefficiency but only production through the technology. \citeasnoun{KARANKI_LIM:2024} estimated a DiD output distance function where the policy shock was the COVID-19 pandemic. They stated they were interested in ``\dots the alterations in technical efficiency caused by the pandemic.'' yet their causal SFM did not directly allow the COVID-19 pandemic to influence technical efficiency.\footnote{We note that the insights of \citeasnoun{GOODMAN-BACON_MARCUS:2020} and \citeasnoun{CALLAWAY_LI:2023} need to be accounted for when using COVID-19 as the policy shock.}

What is needed in the DiD setting is to model the stochastic frontier setup in a single stage which accounts for the before/after, treatment/control grouping, similar to how causal SFA approaches have used RDD \citeaffixed{JOHNES_TSIONAS:2019}{see}. This approach directly treats the variables which distinguish the discontinuity in the error structure and/or the production structure of the firms. The benefit of this approach is that the nature of the RDD is explicitly modeled in a way that is consistent with the notion of a policy discontinuity. 

So in this case we have a model that allows for heterogeneous treatment effects only through the production technology. This would essentially be the model of \citeasnoun{CHEN_ETAL:2020} equation (2.13) but with $u$ not depending upon treatment status. If we also wanted inefficiency to change we would need to have even more care with the structure of the error term. There is also the recent work of \citeasnoun{HENDERSON:2022} that investigates estimation of treatment effects for randomized controlled trials (RCTs) in development contexts where ``inefficiency'' enters as a distinction between functioning of the treatment and an expansion of opportunities (or capabilities) due to the treatment.\footnote{This approach is similar in spirit to the one-sided measurement error model of \citeasnoun{MILLIMET_PARMETER:2022}.}

Here we discuss appropriate statistical approaches to modeling endogeneity in the stochastic frontier model. We first provide a review of standard instrumental variables approaches. IVs are a well-established approach to causal inference in the social sciences, and an important method to add to the toolkit of any causal inference expert (\citename{CUNNINGHAM:2021} \citeyear*{CUNNINGHAM:2021}, Ch. 7, and \citename{KLEIN:2021} \citeyear*{KLEIN:2021}, Ch. 19). We then pivot and discuss the DiD in a stochastic frontier framework. We begin with the simple setting of a policy shock that hits all firms at the same time (i.e. there is no staggering). We discuss what it means to have parallel trends in the setting where both the frontier itself may be impacted by the shock as well as the distribution of inefficiency. This is a subtle but important point. Typically, in the standard DiD setting the error term is assumed to be exogenous with respect to the policy change under study. Here, the policy change is driving changes to inefficiency, which resides in the error term. This means that looking for parallel trends to justify the use of DiD will not work the same as in the regression setting. 

From here we move to discuss the case of staggered DiD and the practical challenges that will arise for practitioners. Given that the error term is one of the key pieces of our focus, the simple two-way fixed effects setup that dominates the applied DiD literature \cite{SUN_ABRAHAM:2021,CALLAWAY_SANTANA:2021} does not map directly to the stochastic frontier model. The error term is a complicated amalgam of different pieces and so the construction of the likelihood function is non-trivial. 

Finally, we provide several remarks on the RDD approach and how it maps into the stochastic frontier setting. As it turns out the RDD approach has been (relatively) more studied and so there does exist a small literature discussing its use. We focus our remarks on the general setup of the model as well as identification issues that are likely to arise from the basic framework. 

One may question if the discussion here is purely a theoretical novelty -- should we worry about endogeneity in frontier settings? That endogeneity matters in the stochastic frontier setting is not a controversial position to take and in fact the literature at large has begun paying more attention to this issue, as will be evidenced in our discussion of instrumental variable approaches. A variety of reasons could lead to the appearance of endogeneity in a frontier setting. It could be that the inputs a firm chooses (in the production frontier) are choice variables, designed to optimize an objective function. Alternatively, omitted variables (such as the level of management) or inputs measured with error (such as capital) could lead to endogeneity. Regardless of its source, the presence of endogeneity will produce inconsistent coefficient estimates and this is doubly troubling here -- our estimates of the production structure, such as returns to scale, are then likely biased, and our estimates of inefficiency are likely to be corrupted. So, developing solutions to handle likely endogeneity in frontier settings is of first order importance.

One popular causal inference tool that we do not develop in detail is the synthetic control method (SCM) of \citeasnoun{Abadie2021JEL}. While SCM has become standard in settings with few treated units and a long pre-treatment panel, its extension to SFA raises additional challenges. In a standard regression framework, SCM constructs a weighted average of control units that matches the pre-treatment trajectory of the treated unit's \emph{observed outcome}. In the SFA framework, however, treatment may affect both the production frontier and the distribution of inefficiency. As such, while SCM can in principle be used to recover a reduced-form treatment effect on observed output, difficulties arise when the researcher seeks to decompose that effect into frontier and inefficiency channels. In particular, such a decomposition would require the synthetic control to match not only the observed output path but also the pre-treatment trajectory of the \emph{unobservable} inefficiency component $u_{it}$. Since $u_{it}$ is latent and can only be estimated conditional on the model parameters, constructing synthetic controls for this purpose would generally require preliminary SFA estimation. Moreover, the standard SCM matching criterion does not distinguish between frontier differences and efficiency differences, so absent additional structure the resulting treatment effect estimate will typically conflate the two channels. We therefore view the extension of SCM to decompositional SFA settings as an important open problem and confine our attention to IV, DiD, and RDD, which map more naturally into the composed-error structure.

Throughout this chapter we illustrate the methods using a setting drawn from \citeasnoun{KUMBHAKAR_TSIONAS_SIPILAINEN:2009}: the choice between organic and conventional dairy farming in Finland. During the period 1995--2002, a number of Finnish dairy farms switched from conventional to organic production methods. This technology choice may affect farm performance through two channels: (a)~a \emph{direct} (frontier) channel, because organic and conventional production technologies differ -- organic farming restricts synthetic inputs and requires different management practices, shifting the production frontier; and (b)~an \emph{indirect} (inefficiency) channel, because the transition to a new technology requires learning, and farmers who switch may initially operate further from their frontier. A na\"ive comparison of output across organic and conventional farms confounds these two channels with selection: farms that choose to adopt organic methods may be systematically different in their efficiency. \citeasnoun{KUMBHAKAR_TSIONAS_SIPILAINEN:2009} address this by jointly estimating the production technologies and the adoption equation using full-information maximum likelihood. In what follows, we show how the same decomposition problem can be addressed using the causal inference toolkit -- IV, DiD, and RDD -- depending on the source of identifying variation available to the researcher. Several common but flawed approaches serve as cautionary benchmarks: (i)~estimating a standard stochastic frontier model ignoring treatment status and then regressing recovered efficiency scores on the adoption indicator (the classic two-step error of \citename{WANG_SCHMIDT:2002} \citeyear*{WANG_SCHMIDT:2002}); (ii)~estimating a probit for adoption and using the inverse Mills ratio as a control in the SFA (which is potentially problematic because the adoption equation itself depends on the unobserved inefficiency); and (iii)~running a standard DiD regression of output on treatment and time dummies, which captures the total effect but cannot separate the frontier shift from the efficiency change. The correct approach embeds both channels in a single-stage model.

\section{Instrumental Variables Approaches}

Before turning to quasi-experimental causal designs, it is useful to briefly review the principal instrumental-variable-based approaches developed to address endogeneity in stochastic frontier settings. These methods represent the natural starting point for causal identification in SFA and provide important context for understanding the appeal of more recent DiD and RDD frameworks. As will be seen throughout this chapter, and as noted by \citeasnoun{AMSLER_PROKHOROV_SCHMIDT:2016}, the composite-error structure of the stochastic frontier model complicates the direct application of standard instrumental-variable and causal inference methods, requiring additional care in implementation.

Begin with the benchmark stochastic frontier model
\begin{equation}\label{eq:sfa}
Y_i=m(X_i;\beta)+v_i-u_i,
\end{equation}
with $Y_i$ output, $m(\cdot;\cdot)$ the parametric production frontier, $X_i$ a $k\times1$ vector of inputs, $v_i$ two-sided noise and $u_i\ge0$ technical inefficiency. To discuss how to incorporate endogeneity into the model write $m(X_i;\beta)=\beta_0+X_{1i}^\prime\beta_1+X_{2i}^\prime\beta_2$ where $X_1$ (of dimension $k_1\times1$) are exogenous inputs and $X_2$ (of dimension $k_2\times1$) are the endogenous inputs -- with $k_1+k_2=k$. Endogeneity may arise through correlation of $X_2$ with $u$, $v$ or both. To focus on the instrumental variable approach, we require instruments $W$ of dimension $\ell\times 1$ with $\ell\ge k_2$. The natural assumption for valid instrumentation is that $W$ is independent of both $u$ and $v$.

\subsection{A Corrected Two Stage Least Squares Approach} 

The most direct approach to handle endogeneity in the stochastic frontier setting is to use corrected ordinary least squares (COLS), but with instruments, what \citeasnoun{AMSLER_PROKHOROV_SCHMIDT:2016} refer to as corrected two stage least squares (C2SLS). This approach estimates the SFM ignoring the error structure using 2SLS with instruments $W$, which produces consistent estimators for $\beta_1$ and $\beta_2$ but not $\beta_0$ (since this is shifted by the unknown $E[u]$ to ensure that the residuals have mean zero). From here, as with standard COLS, higher order moments of the 2SLS residuals are used to solve out for the distributional parameters. Typically only the second and third moments are needed if one is assuming that $u$ stems from a one parameter family (such as Half-Normal or Exponential). Aside from running 2SLS in the first stage, C2SLS is identical to COLS.\footnote{See \citeasnoun{PARMETER:2023} for full details of the COLS approach.} Though the most straightforward, C2SLS does not provide valid standard errors for the intercept or the variance parameters. 

\subsection{A Likelihood Approach}

There are several existing variants of maximum likelihood estimation of the stochastic frontier model \cite{KUTLU:2010,KARAKAPLAN_KUTLU:2017,AMSLER_PROKHOROV_SCHMIDT:2016,CENTORRINO_PEREZ:2019,CENTORRINO_PEREZ_URETA_WALL:2024}. While all conceptually similar, we follow \citeasnoun{AMSLER_PROKHOROV_SCHMIDT:2016} and \citeasnoun{CENTORRINO_PEREZ:2019} as their derivation of the likelihood relies on a simple conditioning argument which we find to be intuitive. 

\citeasnoun{AMSLER_PROKHOROV_SCHMIDT:2016} work with the linear parametric stochastic frontier system:
\begin{align*}\label{eq:SFSys}
Y_i=&m(X_i;\beta)+v_i-u_i=\beta_0+X_{1i}\beta_1+X_{2i}\beta_2+\varepsilon_i\\
X_{2i}=&W_i\Gamma + X_{1i} \Delta  + \eta_i
\end{align*}
and endogeneity of $X_{2i}$ arises through $cov(v_i,\eta_i)\ne0$ (note that endogeneity does not arise through links with inefficiency in this model). \citeasnoun{AMSLER_PROKHOROV_SCHMIDT:2016} assume that $u_i\sim N_+(0,\sigma_u^2)$ and, conditional on $( X_{1i},W_i)$, $\psi_i=(v_i, \eta_i)^\prime\sim N(0,\Omega)$, where
\begin{equation*}
\Omega=\left[\begin{array}{cc}\sigma^2_v&\Sigma_{v\eta}\\\Sigma_{\eta v}& \Sigma_{\eta\eta}\end{array}\right].
\end{equation*}

The likelihood function is derived by conditioning on all the exogenous variables, which yields $f(Y,X_2|X_1,W)=f(Y|X,W) f(X_2|X_1,W)$. With the density in this form, the log-likelihood is: 
\begin{equation*}
\ln \mathcal{L}=\ln\mathcal{L}_1+\ln\mathcal{L}_2,
\end{equation*}
where $\ln\mathcal{L}_1$ corresponds to the likelihood of $f(Y|X,W)$ and $\ln\mathcal{L}_2$ corresponds to the likelihood of $f(X_2|X_1,W)$. Under the maintained distributional assumptions, these components can be written as
\begin{align*}
\ln \mathcal{L}_1=&-(n/2)\ln \sigma^2-\frac{1}{2\sigma^2}\sum\limits^n_{i=1}\tilde\varepsilon_i^2+\sum\limits^n_{i=1}\ln\left[\Phi\left(-\lambda_c\tilde\varepsilon_i/\sigma\right)\right]\\
\ln\mathcal{L}_2=&-(n/2)\ln|\Sigma_{\eta\eta}|-0.5\sum\limits^n_{i=1}\eta_i^\prime\Sigma_{\eta\eta}^{-1} \eta_i,
\end{align*}
where $\tilde\varepsilon_i=Y_i-\beta_0-X_i\beta-\mu_{ci}$, $\mu_{ci}=\Sigma_{v\eta}\Sigma^{-1}_{\eta\eta} \eta_i$, $\sigma^2=\sigma^2_v+\sigma^2_u$, $\lambda_c=\sigma_u/\sigma_c$ and $\sigma^2_c=\sigma^2_v-\Sigma_{v\eta}\Sigma^{-1}_{\eta\eta}\Sigma_{\eta v}$. The subtraction of $\mu_{ci}$ from the common residual, $\varepsilon_i=Y_i-\beta_0-X_i\beta$ in $\ln\mathcal{L}_1$ is an endogeneity correction. It should also be highlighted that $\ln \mathcal{L}_2$ is nothing more than the standard likelihood function of a Multivariate Normal regression model. Estimates of all of the model parameters $(\beta,\sigma^2_v,\sigma^2_u,\Gamma,\Sigma_{v\eta})$ and $\Sigma_{\eta\eta}$ can be obtained through direct maximization of the likelihood function $\ln \mathcal{L}$.

\citeasnoun{CENTORRINO_PEREZ:2019} extend the framework above to allow dependence between the first stage error term, $\eta_i$, and stochastic inefficiency, $u_i$. The idea is to model the distribution of  $u_i \vert \eta_i$ in such a way that, when there is no dependence between $\eta_i$ and $u_i$, the model collapses to the one proposed by \citeasnoun{AMSLER_PROKHOROV_SCHMIDT:2016}. To achieve this, \citeasnoun{CENTORRINO_PEREZ:2019} propose to use a Folded-Normal distribution for $u_i$ with density
\begin{equation} \label{eq:densuclosed}
\begin{aligned}
f(u \vert \eta) =& \frac{1}{\sqrt{2\pi \left( \sigma_u^2 - \Sigma_{u\eta}\Sigma^{-1}_{\eta\eta}\Sigma_{\eta u}\right)}} \left\lbrace \exp \left( -\frac{(u -  \Sigma_{u\eta}\Sigma^{-1}_{\eta\eta}\eta_i)^2}{2(\sigma_U^2- \Sigma_{u\eta}\Sigma^{-1}_{\eta\eta}\Sigma_{\eta u})}\right) \right. \\
&\quad \left. + \exp \left( -\frac{(u +  \Sigma_{u\eta}\Sigma^{-1}_{\eta\eta}\eta)^2}{2(\sigma_U^2 - \Sigma_{u\eta}\Sigma^{-1}_{\eta\eta}\Sigma_{\eta u})}\right)\right\rbrace.
\end{aligned}
\end{equation}
When the vector of dependence parameters, $\Sigma_{\eta u}$, is equal to 0, the density of $u$ is independent of $\eta$, and it becomes the one of a Half-Normal distribution. In the general case, the likelihood function becomes more complex than written above, because of the nonlinearities introduced by the Folded-Normal distribution. However, \citeasnoun{CENTORRINO_PEREZ:2019} show that $\varepsilon_i |\eta_i$ follows a conditional Extended Skew-Normal distributions \cite[p. 35-36]{AZZALINI:2014}, so that the likelihood function can be written in closed form. \citeasnoun{CENTORRINO_PEREZ:2019} leverage the scaling property to extend this framework to the case where there are endogenous environmental variables \citeaffixed[who introduce endogenous environmental variables using a copula approach]{AMSLER_PROKHOROV_SCHMIDT:2017}{see also}.

\subsection{A Method of Moments Approach}

Finally, another novel approach exists that is portable and easy to modify for a variety of distributions for inefficiency. It leverages the corresponding first order conditions in a method of moments approach following \citeasnoun{HANSEN_DONALD_NEWEY:2010}.\footnote{See also the work of \citeasnoun{TRAN_TSIONAS:2013}.} The idea is to use the first order conditions for maximization of the likelihood function under exogeneity:
\begin{align}
&E\left[\varepsilon^2/\sigma^2-1\right]=0 \label{eq:gmm1}\\
&E\left[\frac{\varepsilon\phi_{\lambda,\sigma}(\varepsilon)}{1-\Phi_{\lambda,\sigma}(\varepsilon)}\right]=0 \label{eq:gmm2}\\
&E\left[X\varepsilon/\sigma+\lambda X\frac{\phi_{\lambda,\sigma}(\varepsilon)}{1-\Phi_{\lambda,\sigma}(\varepsilon)}\right]=0,\label{eq:gmm3}
\end{align}
where $\phi_{\lambda,\sigma}(\varepsilon)=\phi(\frac{\lambda\varepsilon}{\sigma})$ and $\Phi_{\lambda,\sigma}(\varepsilon)=\Phi(\frac{\lambda\varepsilon}{\sigma})$. Note that these expectations are taken over $X_i$ and $Y_i$ (and by default, $\varepsilon_i$) and solved for the parameters of the stochastic frontier model.

The key here is that these first order conditions (one for $\sigma^2$, one for $\lambda$ and the vector for $\beta$) are valid under exogeneity and this implies that the maximum likelihood estimator is the generalized methods of moments estimator. Under endogeneity however, this relationship does not hold directly. But the seminal idea of \citeasnoun{AMSLER_PROKHOROV_SCHMIDT:2016} is that the first order conditions \eqref{eq:gmm1} and \eqref{eq:gmm2} are based on the distributional assumptions on $v$ and $u$, not on the relationship of $X$ with $v$ and/or $u$. Thus, these moment conditions are valid whether $X$ contains endogenous components or not. The only moment condition that needs to be adjusted is \eqref{eq:gmm3}. In this case the first order conditions need to be taken with respect to $W$, the exogenous variables, not $X$. Doing so results in the following amended first order conditions:
\begin{equation}
E\left[W_i\varepsilon_i/\sigma+\lambda W_i\frac{\phi_i}{1-\Phi_i}\right]=0,\label{eq:gmm4}
\end{equation}
where $\phi_i$ and $\Phi_i$ are identical to those in \eqref{eq:gmm3}. It is important to acknowledge that these moment conditions are valid when $\varepsilon_i$ and $W_i$ are independent. This is a more stringent requirement than the typical regression setup with $E[\varepsilon_i|W_i]=0$. As with the C2SLS approach, the source of endogeneity for $X_2$ does not need to be specified (through $v$ and/or $u$).

The theoretical frameworks above have begun to find empirical traction. \citeasnoun{KARAKAPLAN_KUTLU:2017} apply the endogenous panel SFA model to the Japanese cotton spinning industry, demonstrating that ignoring the endogeneity of inputs leads to substantial bias in efficiency estimates. \citeasnoun{CENTORRINO_PEREZ:2019} propose a maximum likelihood estimator for SFA with endogeneity based on control functions and a Folded-Normal specification for the conditional distribution of inefficiency; their application to rice farmers in Nepal shows that accounting for the endogeneity of inputs substantially affects both frontier parameter estimates and the ranking of individual efficiency scores. \citeasnoun{CENTORRINO_PEREZ_URETA_WALL:2024} extend this framework to the case of an endogenous binary treatment, studying a soil conservation program in El Salvador; their model allows both the frontier and inefficiency to depend on a potentially endogenous participation indicator, with instrumental variables used to define the assignment mechanism. Applied researchers can also leverage the Stata command \texttt{xtsfkk} \citeaffixed{Karakaplan2022}{see}, which implements panel SFA with endogenous regressors and endogenous inefficiency.

An instrument-free alternative is the approach of \citeasnoun{TRAN2015} and \citeasnoun{Haschka2024}, who model the dependence between endogenous regressors and the composed error via a copula. This approach is particularly useful when external instruments are weak or unavailable, and it can also handle cases where the residual skewness contradicts standard SFA assumptions (the ``wrong skewness'' problem). Identification rests on the copula and distributional shape rather than on exclusion restrictions.\footnote{A distinct but related form of endogeneity arises from sample selection: if program participation correlates with latent inefficiency, the observed sample is not representative. \citeasnoun{Greene2010} extends the Heckman selection correction to the SFA setting, providing a natural first step when selection into treatment is plausibly non-random.}\footnote{\citeasnoun{TSIONAS_IZZELDIN_HENNINGSEN_PARAVALOS:2022} develop a Bayesian approach to addressing endogeneity when estimating stochastic ray production frontiers, offering full posterior inference on efficiency distributions. This Bayesian route is attractive when the classical likelihood is complex or when distributional uncertainty on individual efficiency estimates is of interest.}

When panel data are available and the source of endogeneity is time-invariant heterogeneity correlated with regressors, correlated random-effects (CRE) SFA models \citeaffixed{KaragiannisKellermann2019,HajargashtGriffiths2019}{see} provide a parsimonious alternative to IV. In the Mundlak/Chamberlain tradition, the conditional mean of $u_i$ (or $v_i$) is specified as a function of group means of the regressors, absorbing the within-group correlation without requiring external instruments. This path is especially valuable alongside DiD/RDD when the researcher cannot credibly exclude time-invariant confounders.

This is a brief review of the extant literature on endogeneity in the stochastic frontier model.\footnote{We refer to \citeasnoun{KUMBHAKAR_PARMETER_ZELENYUK:2021} for a more detailed overview and additional discussion.} The key point for the remainder of this chapter is that, while these approaches provide powerful tools when credible instruments are available, their practical implementation often hinges on strong exclusion restrictions and valid instrument selection. In many empirical applications such assumptions are difficult to justify, and the methods themselves offer limited guidance regarding how instruments should be chosen. As a result, researchers may instead seek to exploit quasi-random variation arising from institutional or policy settings. This has led to growing interest within the productivity and efficiency literature in quasi-experimental designs such as DiD and RDD. We therefore now turn to these modern causal inference approaches in the stochastic frontier setting.

\section{Overview -- Perfect Random Assignment}

Consider a potential outcomes setup following \citeasnoun{ANGRIST_PISCHKE:2014}: $Y_i(0)$ is the outcome in which farm $i$ does not receive treatment (say a farm continues with conventional production) and $Y_i(1)$ is the outcome if farm $i$ does receive the treatment (say the farm adopts organic technology). If we let $D_i$ be a dummy variable which captures treatment assignment, then the $i$th observed outcome is $Y_i=Y_i(0)(1-D_i)+Y_i(1)D_i$. When treatment is randomly assigned the causal effect is $\tau=Y_i(1)-Y_i(0)$, but this parameter is unobserved since we never observe both states of the world. 

A better option is to focus on the average treatment effect, $\tau_{ATE}=E[Y_i(1)|D_i=1]-E[Y_i(0)|D_i=0]$. Assuming a constant treatment effect across all farms, $\tau_{ATE}$ is a good measure for $\tau$. Naturally if there are heterogeneous effects then $\tau_{ATE}$ is not expected to be a good representation.

Now suppose we have the following setup
\begin{align*}
Y_i(0)=&m(0)+v_0\\
Y_i(1)=&m(1)+v_1.
\end{align*}
A common (needed) assumption is that $E[v_1|D]=E[v_0|D]=0$ and, for a constant treatment effect, we assume $m(1)=m(0)+\tau$. In this case $E[Y_i|D_i=1]-E[Y_i|D_i=0]$ is a valid estimator of the average treatment effect. But suppose we have technical efficiency. If we have a policy that is designed to both improve output directly through technical change and indirectly through efficiency change then our average treatment effect looks much different. 

In this setup we have 
\begin{align*}
Y_i(0)=&m(0)+v_0-u_0=m(0)+\varepsilon_0\\
Y_i(1)=&m(1)+v_1-u_1=m(1)+\varepsilon_1.
\end{align*}
Here we assume $E[\varepsilon_1|D]\ne E[\varepsilon_0|D]\ne0$. Further 
\begin{align*}
E[Y_i|D_i=1]-E[Y_i|D_i=0]=&m(1)-m(0)+E[\varepsilon_{1i}|D_i=1]-E[\varepsilon_{0i}|D_i=0]\\=&m(1)-m(0)-(E[u_{1i}|D_i=1]-E[u_{0i}|D_i=0])\\=&\tau-(E[u_{1i}|D_i=1]-E[u_{0i}|D_i=0]).
\end{align*}
Thus, in the presence of inefficiency, unless a treatment has no effect on inefficiency, the average treatment effect measures the causal effect minus the change in efficiency. Again, assuming random assignment the ATE could easily be estimated by dividing the sample into control ($D_i=0$) and treatment ($D_i=1$) groups and taking the difference in average $Y_i$. As just noted, this does not allow separation of the competing effects. Rather, one needs to decompose them into direct effects on technology and indirect effects through inefficiency.

\begin{definition}[Direct and Indirect Treatment Effects]\label{def:direct-indirect}
For a unit $i$, the \emph{direct} (frontier) treatment effect is the change in expected output attributable to a shift in the production technology:
\[
\mathrm{ATE}_{Tech} = m(1)-m(0).
\]
The \emph{indirect} (inefficiency) treatment effect is the change in expected inefficiency:
\[
\mathrm{ATE}_{Ineff} = E[u_1|D=1]-E[u_0|D=0].
\]
The observed average treatment effect satisfies $\mathrm{ATE} = \mathrm{ATE}_{Tech} - \mathrm{ATE}_{Ineff}$.
\end{definition}

In a simple cross-sectional setting separate estimation of the direct and indirect effect could be handled by simply modeling the two different inefficiency distributions. For example, if we assume that $v_1\sim v_0\sim N(0,\sigma_v^2)$, $u_1\sim N_+(0,\sigma_{u1}^2)$ and $u_0\sim N_+(0,\sigma_{u0}^2)$, then the composed error has the classic Normal-Half Normal shape but with differing pretruncation variance parameters. That is, for $D_i=0$, the density is 
\begin{equation}
f_0(\varepsilon_i)=\frac{2}{\sigma_0}\phi(\varepsilon_i/\sigma_0)\Phi(\lambda_0\varepsilon_i/\sigma_0)
\end{equation}
and 
for $D_i=1$, the density is 
\begin{equation}
f_1(\varepsilon_i)=\frac{2}{\sigma_1}\phi(\varepsilon_i/\sigma_1)\Phi(\lambda_1\varepsilon_i/\sigma_1),
\end{equation}
where $\sigma_j^2=\sigma_v^2+\sigma_{uj}^2$ and $\lambda_j=\sigma_{uj}/\sigma_v$ for $j=0,1$ and $\varepsilon_i=Y_i-\alpha- D_i\tau$ where we have condensed the notation $m(0)=\alpha$ and $m(1)=\alpha+\tau$. If we model the variance of the inefficiency component we begin by writing $\sigma_{u0}=e^{\gamma_0}$ and $\sigma_{u1}=e^{\gamma_1}$ which can be rewritten as $\sigma_{u1}=e^{\gamma_0+D_i\gamma_1}=\sigma_{u0}e^{D_i\gamma_1}$. Now we may write our likelihood function as 
\begin{equation}
\mathcal{L}=\prod\limits^n_{i=1}f_0(\varepsilon_i)^{1\{D_i=0\}}f_1(\varepsilon_i)^{1\{D_i=1\}}
\end{equation}
which upon taking logarithms produces
\begin{align}
\ln\mathcal{L}=&\sum\limits^n_{i=1}\left[1\{D_i=0\}\ln f_0(\varepsilon_i)+1\{D_i=1\}\ln f_1(\varepsilon_i)\right]\notag\\=&\sum\limits^{n_0}_{i=1}\ell^0_{i}+\sum\limits^{n_1}_{i=1}\ell^1_{i},
\end{align}
where $\ell^j_i$ are the log-likelihood contributions for $j\in\{0,1\}$ of the control and treated groups, $n_0$ is the number of farms in the control group and $n_1$ is the number of farms in the treated group. More specifically we have that 
\begin{equation}
\ell^0_{i}=-\ln\sigma_0-\frac{1}{2\sigma^2_0}\left(\varepsilon^0_i\right)^2+\ln\Phi\left(-\lambda_0\varepsilon^0_i/\sigma_0\right)
\end{equation}
and 
\begin{equation}
\ell^1_{i}=-\ln\sigma_1-\frac{1}{2\sigma^2_1}\left(\varepsilon^1_i\right)^2+\ln\Phi\left(-\lambda_1\varepsilon^1_i/\sigma_1\right)
\end{equation}
where $\varepsilon^0_i=Y_i-\alpha$ and $\varepsilon^1_i=Y_i-\alpha-\tau$.

Outside of the specific dummy variable capturing treatment status, this is just a pooled cross-section maximum likelihood problem. The treated and control observations are pooled together and a dummy variable is introduced for those units that are treated to help distinguish the direct and indirect effects. Further, the `complication' is that both the treated and control groups share the same noise variance, $\sigma_v$. If $n$ were large and efficiency concerns ignored one could simply split the sample, estimate two separate stochastic frontier models and make the necessary calculations to determine the average treatment effect. Unfortunately, while this setup is simple, it is highly unrealistic in practice since it is unlikely to mimic practical situations (among other empirical issues). 

This model can be easily adapted to allow for covariates and to allow the effect of those covariates to differ between treatment and control. That is, if the technology changes by assignment status, then, for linear in parameters technology $m(X;\beta)=X^\prime\beta$, we would have $m_0(X;\beta_0)=X^\prime\beta_0$ and $m_1(X;\beta)=X^\prime\beta_1$ and our errors now become $\varepsilon_i=Y_i-\alpha-\tau D_i-(X^\prime\beta_0) (1-D_i)-(X^\prime\beta_1) D_i$. In this setup we assume that the technology changes by a constant amount, i.e. the policy is factor neutral. One can easily manipulate the model to allow a policy to be factor non-neutral if desired. 

An alternative to using maximum likelihood would be to instead deploy COLS (\citename{OLSON_SCHMIDT_WALDMAN:1980} \citeyear*{OLSON_SCHMIDT_WALDMAN:1980} and \citename{PARMETER:2023} \citeyear*{PARMETER:2023}). In this case we can recover the dual treatment effects through linear regression and moment matching. First, the direct effect is estimated running the regression of $Y$ on a constant and the dummy variable capturing treatment assignment. Second, the residuals from this regression are then used to solve two moment equations based on the distributional choices made for $v$ and $u$. In this case we need to solve the moment equations such that the constancy of $\sigma_v$ is preserved across the groups. 

Denote the residuals as $\widehat{\varepsilon}_i$. Then we have $Var(\varepsilon|D=j)=\sigma_v^2+\sigma^2_{uj}(1-2/\pi)=\sigma_v^2+c_2\sigma^2_{uj}$ and the third moment is $\mu_3(\varepsilon|D=j)=-\mu_3(u_j)=\frac{\sqrt{2}(4-\pi)}{\pi^{3/2}}\sigma^3_{uj}=c_3\sigma^3_{uj}$. Then the moment matching produces the estimators 
\begin{equation}
\widehat\sigma_{uj}=\left(\frac{-m_{3j}}{c_3}\right)^{1/3},\quad \widehat{\sigma}^2_v=\sum\limits_{j\in\{0,1\}}\frac{n_j}{n_0+n_1}\left(m_{2j}-c_2\widehat\sigma^2_{uj}\right),
\end{equation}
where the empirical moments are calculated as 
\begin{equation}
m_{2j}=n_j^{-1}\sum\limits_{i\in\{D_i=j\}}\left(\widehat\varepsilon_i-\bar{\widehat\varepsilon}_i\right)^2,\quad m_{3j}=n_j^{-1}\sum\limits_{i\in\{D_i=j\}}\left(\widehat\varepsilon_i-\bar{\widehat\varepsilon}_i\right)^3.
\end{equation}

\section{Equal Timing}

Now suppose that we do not have random assignment. It is clear that the framework described above will not work. We discuss here the traditional DiD framework, but within the context of the stochastic frontier model. To begin, assume there are two periods, $T = \left\lbrace 0 , 1 \right\rbrace$. In period $0$ no units are treated and in period $1$ some of the units are treated (in our running example, a group of conventional dairy farms switch to organic production). The classic DiD with no covariates would calculate the average treatment effect by 
\begin{equation}
\tau_{DiD}=\left(E[Y(1)|T = 1]-E[Y(1)|T = 0]\right)-\left(E[Y(0)|T = 1]-E[Y(0)|T = 0]\right).
\end{equation}
In this setting we cannot observe all of the outcomes since only some of the units are treated, and only in period $1$. So we look at the difference in means of those farms before and after the treatment who were treated and compare it to the difference in means of those farms before and after the treatment who were not treated:
\begin{equation}
\widehat{\tau}_{DiD}=\left(\bar{Y}_{11}-\bar{Y}_{10}\right)-\left(\bar{Y}_{01}-\bar{Y}_{00}\right),
\end{equation}
where $\bar{Y}_{00}=n_{00}^{-1}\sum\limits^{n_{00}}_{k=1}Y_{k}$, $\bar{Y}_{10}=n_{10}^{-1}\sum\limits^{n_{10}}_{k=1}Y_{k}$, $\bar{Y}_{01}=n_{01}^{-1}\sum\limits^{n_{01}}_{k=1}Y_{k}$ and $\bar{Y}_{11}=n_{11}^{-1}\sum\limits^{n_{11}}_{k=1}Y_{k}$, where $n_{ij}$ is the number of observations such that $D = i$ and $T = j$, with $i,j = \left\lbrace 0,1 \right\rbrace$. In a regression framework the DiD estimator can be obtained from 
\begin{equation} \label{eq:didregspec}
Y_i=\beta_0+D_i\beta_1+T_i\beta_2+D_i T_i \beta_3+\varepsilon_i
\end{equation}
where $\widehat\beta_3\equiv\widehat{\tau}_{DiD}$. 

Now, to understand the complication of constructing a DiD in the stochastic frontier setting, again, consider that the mean of $Y$ can be influenced directly through the policy (as is assumed in the traditional DiD construction), but also indirectly through the policy via changes in inefficiency. So if we write out our DiD formulation, we have 
\begin{equation}
\tau_{DiD}=\left(E[Y(1)|T = 1]-E[Y(0)|T = 1]\right)-\left(E[Y(1)|T=0]-E[Y(0)|T = 0]\right).
\end{equation}
The key here is to recognize that this is just repeating the discussion above with random assignment twice, once in period $1$ and once in period $0$. In turn, the lack of random assignment can be overcome in the stochastic frontier setting by looking at assignment through the changed induced by the policy over time. 

We compare this setup to two alternative approaches that could be deployed. The first approach would estimate the stochastic production frontier model, ignoring assignment, calculate technical inefficiencies and then run a DiD regression using these estimated inefficiencies. This approach is generally unattractive. While conceptually pleasant to describe and straightforward to implement empirically, it does not estimate the treatment effect and suffers from a host of problems related to the `two-step' approach as detailed in \citeasnoun{WANG_SCHMIDT:2002}. 

The second approach would construct a DiD type variable setup and include this in the parameter(s) of the inefficiency distribution and then estimate the stochastic frontier model via maximum likelihood analysis as per usual. This approach, when properly implemented, can provide an accurate estimate of the direct and indirect effects of the treatment upon additional assumptions. For instance, using the same specification as in equation \eqref{eq:didregspec} and augmenting it to include inefficiency, we can write
\begin{equation}
Y_i=\beta_0+D_i\beta_1+T_i\beta_2+D_i T_i\beta_3+v_i - u_{0i} e^{D_i\gamma_1+T_i\gamma_2+D_i T_i\gamma_3},
\end{equation}
where $v_i$ is a two sided error term which satisfies $E \left[ v_i \vert D_i,T_i \right] =  E \left[ v^3_{i} \vert D_i,T_i \right] = 0$, $u_{0i} \sim N^+ (0,\sigma^2_u)$, and $v_i$ and $u_{0i}$ are independent. Note that in this DiD setting, the decomposition of Definition~\ref{def:direct-indirect} naturally extends to the average treatment effect on the treated, conditioning on $D_i=1$. Then, the following moment conditions hold
\begin{align*}
E\left[ Y_i \vert D_i = 0, T_i  = 0 \right]  =& \beta_0  - E\left[ u_{0i}\right] = \beta_0 - \sqrt{\frac{2}{\pi}} \sigma_u\\ 
E\left[ Y_i \vert D_i = 1, T_i  = 0 \right]  =& \beta_0  + \beta_1  - \sqrt{\frac{2}{\pi}} \sigma_u e^{\gamma_1} \\
E\left[ Y_i \vert D_i = 0, T_i  = 1 \right]  =& \beta_0  + \beta_2 - \sqrt{\frac{2}{\pi}} \sigma_u e^{\gamma_2} \\
E\left[ Y_i \vert D_i = 1, T_i  = 1 \right]  =& \beta_0  + \beta_1 + \beta_2 + \beta_3 - \sqrt{\frac{2}{\pi}} \sigma_u e^{\gamma_1 + \gamma_2  + \gamma_3}.
\end{align*}
The identification of $\gamma_1$, $\gamma_2$ and $\gamma_3$ from the four moment conditions above rests on two maintained assumptions: that $v_i$ is Normal, $u_{0i}$ is Half-Normal, and that the scaling form $u_{0i}e^{D_i\gamma_1+T_i\gamma_2+D_iT_i\gamma_3}$ correctly captures how inefficiency responds to treatment and time. These assumptions are standard in applied SFA but not innocuous; relaxing either the distributional assumptions or the scaling form returns us to the open methodological problem discussed in Section \ref{sec:open-parallel}.

We can separately identify $\beta_0$ and $\sigma_u$ using a COLS-type approach. Because of our assumptions about the error term
\begin{align*}
E & \left[ \left(Y_i - E\left[ Y_i \vert D_i = 0, T_i  = 0 \right]\right)^3_i \vert D_i = 0, T_i  = 0 \right]  = - E\left[\left( u_{0i} - E\left[  u_{0i} \right] \right)^3 \right]  \\
& \qquad = - \sigma^3_u \sqrt{\frac{2}{\pi}}\left(\frac{4 - \pi}{\pi} \right),
\end{align*}
which identifies $\sigma_u$. Therefore, the estimand of $\beta_0$ is 
\[
\beta_0  = E\left[ Y_i \vert D_i = 0, T_i  = 0 \right] +  \sqrt{\frac{2}{\pi}} \sigma_u. 
\]
In a similar way, and for given $\beta_0$ and $\sigma_u$, one can identify $\beta_1$ and $\gamma_1$ from the second equation, and so on and so forth. Finally, $\beta_3$ which captures the direct treatment effect, and $\gamma_3$, the indirect treatment effect can be identified from the last equation. A likelihood-ratio test can be used to test whether there are indirect effects of the treatment through inefficiency. 

Extensions to the case where the production frontier and the inefficiency depend on time-invariant covariates should be straightforward to implement. 

For applied researchers, the single-stage likelihood given by the DiD regression specification above can be estimated in Stata using the {\tt xtsfkk} package of \citeasnoun{Karakaplan2022}, which accommodates endogenous regressors and a scaling inefficiency specification. In practice we recommend four steps: (i) test for parallel trends in the reduced-form outcome before imposing the SFA structure, keeping in mind the caveats of Section 7.1; (ii) estimate the single-stage model with the scaling specification and verify that the $\gamma$ parameters are identified (i.e., that the variance of $u$ is separately estimable from the variance of $v$); (iii) use a likelihood-ratio test for the null that indirect effects are zero before interpreting the decomposition; and (iv) check robustness by re-estimating the model under alternative inefficiency distributions (Exponential, Gamma). A complete empirical application following this template is provided by \citeasnoun{BRAVO_URETA_ETAL:2020}, discussed next.

\citeasnoun{BRAVO_URETA_ETAL:2020}, develop a DiD selectivity-corrected stochastic production frontier model to evaluate the PAGRICC environmental program in Nicaragua (2012--2016). Their framework combines propensity score matching with a single-step SFA that separately identifies technology change (a shift in the frontier) and technical efficiency change (a shift in $u$) attributable to the intervention, while controlling for self-selection into treatment using the methodology of \citeasnoun{Greene2010}. They find that while a severe drought had significant negative effects on both treated and control farms, project beneficiaries enjoyed significantly better outcomes -- and that the decomposition into frontier and efficiency components reveals low overall efficiency levels with no significant variation across models, time, or treatment status. This study illustrates how embedding DiD directly in the SFA likelihood avoids the pitfalls of two-step approaches and enables the decomposition of the treatment effect into its direct and indirect components.

\section{Heterogeneous Timing}

We now move to the general heterogeneous timing, heterogeneous effect event study setup, but introduce the model in the stochastic frontier setting. In this case we will need some additional notation. We observe a random sample of $i=1,\ldots,n$ units over $t=0,\ldots,T$ time periods. For each unit we observe $Y_{it}$ (the outcome of interest) and treatment status $D_{it}\in\{0,1\}$. We have the generic $D_{it}=0$ if unit $i$ is untreated in period $t$ and is equal to 1 if the unit is treated. We will focus on the absorbing treatment case. Once a unit is treated, they stay treated through time $T$. The key focus in recovering treatment effects is the period at which the unit switches from being untreated to being treated, the initial treatment, defined as $E_i=\min\{t:D_{it}=1\}$. For a never treated unit $E_i=\infty$. 

As in \citeasnoun{SUN_ABRAHAM:2021}, units can also be uniquely characterized into disjoint cohorts $e$ for $e\in\{0,\ldots,T,\infty\}$ -- we divide all units into cohorts based on when they are treated. That is, units in cohort $e$ are all first treated at the same time $\{i:E_i=e\}$. This effectively gives us two ways to count time, calendar time, where we look at units relative to the calendar (period 1, period 2, etc.) and treatment time, where we look at time since treatment (1 period of being treated, 2 periods of being treated, etc.). In our running example, units in cohort $e$ are Finnish dairy farms that switched from conventional to organic production in calendar year $e$ during 1995--2002, while farms that never switched constitute the never-treated cohort with $E_i=\infty$.

Define $Y_{it}^e$ to be the potential outcome of unit $i$ in period $t$ who is first treated in period $e$. $Y^\infty_{it}$ is defined as the potential outcome for those units that never receive treatment. Thus, the observed outcome can be defined as 
\begin{equation}
Y_{it}=Y^{E_i}_{i,t}=Y^\infty_{it}+\sum\limits_{0\leq e\leq T}\left(Y_{it}^e-Y^\infty_{it}\right)1\{E_i=e\}.
\end{equation}
The unit-level treatment effect is the difference between the observed outcome and the counterfactual outcome of never-being treated: $Y_{it}-Y^\infty_{it}$. For those units that are not treated this is necessarily 0, while for those units that are treated this represents a reasonable counterfactual \cite{SUN_ABRAHAM:2021}. Other counterfactuals could easily be considered. 

Following \citeasnoun{SUN_ABRAHAM:2021}, the cohort-specific average treatment effect on the treated (CATT) $\ell$ periods after initial treatment is 
\begin{equation}
CATT_{e,\ell}=E\left[Y_{ie+\ell}-Y^\infty_{ie+\ell}|E_i=e\right].
\end{equation}
These treatment effects vary based on time from initial treatment ($\ell$) and on initial treatment time ($e$). Thus, $e$ is based on calendar time and $\ell$ is a relative time (since different cohorts can all be $\ell$ periods from initial treatment).\footnote{More specifically, $0\leq e\leq T$ while $-e\leq \ell\leq T-e$. } This allows comparisons of cohorts (which is of interest when treatment is not uniform across the population). 

We maintain the same three assumptions as in \citeasnoun{SUN_ABRAHAM:2021}: (i) parallel trends in baseline outcomes; (ii) no anticipatory behavior prior to treatment; (iii) treatment effect homogeneity. Parallel trends essentially removes time effects from contaminating the CATT. This assumption is quite strong and could be lessened in a variety of ways. No anticipatory behavior rules out changes in behavior prior to treatment which then corrupt estimates of CATT. Finally, treatment effect homogeneity assumes that $CATT_{e,\ell}=CATT_{\ell} = ATT_{\ell}$, so all treated units have the same cohort effects and are invariant to when they are treated. Crucially, the classical parallel-trends assumption, stated in terms of observed outcomes, does not map directly onto the SFA decomposition: it leaves open whether identification requires parallel trends in the frontier component, in the inefficiency component, or only in their composite. We position the SFA-specific formalization of parallel trends as an unresolved methodological problem in Section \ref{sec:open-parallel}.

In a nutshell, in a canonical two-way fixed-effects (TWFE) regression with staggered treatment timing, the coefficient on the treatment indicator is a weighted average of CATTs, where some weights can be \emph{negative}. This occurs because already-treated cohorts are used as implicit controls for late-treated cohorts, contaminating the estimates when treatment effects are heterogeneous across cohorts. In the SFA setting this problem is compounded: the TWFE regression mixes \emph{direct} (frontier) and \emph{indirect} (inefficiency) treatment effects within the same weighted average, making it impossible to sign or interpret the coefficient without additional structure. The interaction-weighted estimator of \citeasnoun{SUN_ABRAHAM:2021} solves the first problem (negative weights) but does not automatically solve the second (direct vs.\ indirect confounding). The decomposition we develop below addresses both issues jointly. For applied researchers seeking additional guidance on staggered DiD, event-study design, and the choice among modern estimators, we recommend \citeasnoun{RothEtal2023} and the practitioner's guide of \citeasnoun{BakerEtal2025}.

\citeasnoun[Eq. 4]{SUN_ABRAHAM:2021} study the properties of the two-way fixed effect regression model
\begin{align}\label{eq:exo_treat}
Y_{it} & =\theta_i+\delta_t+\sum\limits_{g\in\mathcal{G}}\mu_g 1\{t-E_i\in g\}+v_{it}. 
\end{align}
Under the three assumptions above, they show that the parameters $\lbrace \mu_g, g \in \mathcal{G} \rbrace$ only identify a weighted average of the $(C)ATT_\ell$, whose weights are not positive and do not necessarily sum up to one. They therefore propose to obtain an estimator of the $(C)ATT_\ell$ from the following DiD specification
\begin{equation} \label{eq:exo_treat_iw}
Y_{it} = \theta_i + \delta_t + \sum_{e \notin C} \sum_{\ell \neq -1} \delta_{e,\ell} 1\{E_i  = e \} D^\ell_{it} + v_{it},
\end{equation}
where the baseline period is the period right before treatment (period $-1$), and $C$ is the control group. Under a parallel trend assumption only, they show that 
\begin{equation} \label{eq:delta_sa_iw}
\hat{\delta}_{e,\ell} = \frac{\mathbb{E}_n\left[ \left( Y_{i,e + \ell} - Y_{i,-1} \right) 1\{E_i  = e \}\right] }{\mathbb{E}_n \left[ 1\{E_i  = e \}  \right] } -  \frac{\mathbb{E}_n\left[ \left( Y_{i,e + \ell} - Y_{i,-1} \right) 1\{E_i  \in C \}\right] }{\mathbb{E}_n \left[ 1\{E_i  \in C \}  \right] },
\end{equation}
is an unbiased and consistent estimator of $CATT_{e,\ell}$, where $\mathbb{E}_n = \frac{1}{n}\sum_{i = 1}^n$ denotes sample averages.

We now wish to allow the expectation of $Y$ to depend on inefficiency, so that the $CATT$ can be decomposed as follows
\begin{equation}
CATT_{e,\ell}= CATT_{e,\ell,Tech} - CATT_{e,\ell,Ineff}.
\end{equation}

The two-way fixed effect model in equation \eqref{eq:exo_treat} becomes 
\[
Y_{it} =\theta_i+\delta_t+\sum\limits_{g\in\mathcal{G}}\mu_g 1\{t-E_i\in g\}+v_{it} - u_{it}. 
\]
The main negative result in \citeasnoun{SUN_ABRAHAM:2021} remains unaffected by this modification, except that now the TWFE regression identifies a weighted combination of the direct and indirect treatment effects. That is, letting
\begin{align*}
f_{Tech} (e,t) =& CATT_{e,t,Tech}  - CATT_{e,\infty,Tech} \\
f_{Ineff} (e,t) =& CATT_{e,t,Ineff}  - CATT_{e,\infty,Ineff},
\end{align*}
the result in their Proposition 1 can be rewritten as 
\begin{align*}
\mu_g =&  \sum_{\ell \in g} \sum_{e \neq \infty} \omega^g_{e,\ell } \left[ f_{Tech} (e,e + \ell )  - f_{Ineff}(e,e + \ell )  \right] \\
& \qquad + \sum_{g^\prime \neq g, g^\prime \in \mathcal{G}}  \sum_{\ell \in g^\prime} \sum_{e \neq \infty} \omega^g_{e,\ell } \left[ f_{Tech} (e,e + \ell )  - f_{Ineff}(e,e + \ell )  \right] \\
& \qquad + \sum_{\ell \in g^{excl}} \sum_{e \neq \infty} \omega^g_{e,\ell } \left[ f_{Tech} (e,e + \ell )  - f_{Ineff}(e,e + \ell )  \right] \\
& \qquad + \sum_{t}  \omega^g_{\infty,t } \left[ f_{Tech} (\infty,t )  - f_{Ineff}(\infty,t )  \right], 
\end{align*}
where $g^{excl}$ denotes the excluded event-time bin $\{-1\}$ in the notation of \citeasnoun{SUN_ABRAHAM:2021}, and the weights $\omega^g$ can be computed from the same auxiliary regression models. Without further considerations, if we were to assume that $u_{it}$ is independent of the treatment, then $E\left[ u_{i,t} - u^{\infty}_{i0} \vert E_i = e\right] = E\left[ u_{i,t} - u^{\infty}_{i0} \right]$, and $f_{Ineff} (e,t) = 0$, for all $(e,t)$. In this case, we can immediately follow the approach in \citeasnoun{SUN_ABRAHAM:2021}, and obtain an unbiased and consistent estimator of $CATT_{e,\ell}$ from $\hat{\delta}_{e,\ell}$ in equation \eqref{eq:delta_sa_iw} above. The intuition for this result is that the presence of inefficiency would only affect the intercept of the regression, which is captured by the time and farm fixed effects. However, those are removed by taking differences with respect to a control group ($C$), and a baseline period ($-1$), respectively. 

Alternatively, we may also want to allow for indirect effects of the treatment. In this case, the inefficiency distribution changes by cohort and therefore the estimator of $\delta_{e,\ell}$ in this model
\begin{align}\label{eq:exo_treat_ineff}
Y_{it} & = \theta_i + \delta_t + \sum_{e \notin C} \sum_{\ell \neq -1} \delta_{e,\ell} 1\{E_i  = e \} D^\ell_{it} + v_{it}-u_{it}, 
\end{align}
would confound direct and indirect effects. That is,
\begin{align*}
\hat{\delta}_{e,\ell} =& \frac{\mathbb{E}_n\left[ \left( f_{Tech}(E_i, E_i + \ell) - f_{Tech}(E_i,-1) \right) 1\{E_i  = e \}\right] }{\mathbb{E}_n \left[ 1\{E_i  = e \}  \right] } \\
& \qquad  -  \frac{\mathbb{E}_n \left[  \left( f_{Tech}(E_i, E_i + \ell) - f_{Tech}(E_i,-1) \right) 1\{E_i  \in C \}\right] }{\mathbb{E}_n \left[ 1\{E_i  \in C \}  \right] }\\
& \qquad  -  \frac{\mathbb{E}_n\left[ \left( u_{i,e + \ell} - u_{i,-1} \right) 1\{E_i  = e \}\right] }{\mathbb{E}_n \left[ 1\{E_i  = e \}  \right] } + \frac{\mathbb{E}_n\left[ \left( u_{i,e + \ell} - u_{i,-1} \right) 1\{E_i  \in C \}\right] }{\mathbb{E}_n \left[ 1\{E_i  \in C \}  \right] }.
\end{align*}

It is clear that if one ignores the potential effects of a policy change on inefficiency, then the direct estimator proposed in \citeasnoun{SUN_ABRAHAM:2021} is downward biased by the amount of the indirect effect. 

The open issue is under what conditions one can hope to separately recover the direct and indirect effects. A direct maximum likelihood approach on the model in equation \eqref{eq:exo_treat_ineff} is not viable, as it would suffer from the incidental parameter problem \cite{GREENE:2005}. We can potentially follow either \citeasnoun{CHEN_ETAL:2014} or \citeasnoun{BELOTTI_ILARDI:2018} to estimate the model. Under the assumption that $u_{it}$ follows a Half-Normal distribution and $v_{it}$ follows a Normal distribution, \citeasnoun[Eq (18)]{CHEN_ETAL:2014} use the fact that $v_{it}-u_{it}$ is a member of the Skew Normal family and so a within transformation on it produces a new random variable that is a member of the Closed Skew Normal family. This result could be leveraged to construct a maximum likelihood estimator of direct and indirect effects. We defer a comprehensive analysis of this case to further research; the broader open problem of how parallel trends should be formalized when treatment acts on both the frontier and the inefficiency distribution is discussed in Section~\ref{sec:open-parallel}.

\section{Regression Discontinuity Designs}

RDDs also leverage quasi-experimental variation, as with DiD, but exploit a different source of variation. Whereas DiD relies on (quasi-)exogenous assignment into or out of a treatment protocol \textit{over time}, RDDs use assignment relative to a threshold and compare farms on either side of that cutoff that are otherwise similar. For simplicity, and in contrast to the time-based structure of DiD, we develop the RDD framework in a cross-sectional setting. For example, if only farms below a certain threshold size qualify for transitioning to organic farming based on a given asset level $A$, we may compare farms on either side of $A$ that are otherwise similar. In this case, assignment near the cutoff can be viewed as as-good-as-random. A key feature of an RDD is that the threshold variable (the running variable) may be correlated with outcomes, but this relationship is assumed to be smooth, in the sense that small changes in $A$ lead to small changes in $Y(0)$ and $Y(1)$.\footnote{To credibly identify the policy effect, the assignment (running) variable should not be manipulable, and any variation in incentives around the cutoff should be as-if random. In practice, the analyst should (i) verify a discontinuous jump in the outcome—typically via an inspection plot; (ii) check that observable covariates are smooth and similarly distributed around the cutoff; and (iii) test the continuity of the running variable (e.g., a McCrary density test) to rule out manipulation.}

This threshold arises from policy or administrative procedures driven by resource constraints. The allocation of such a covariate is based not on administrator discretion but on well-planned, transparent rules. In \citeasnoun{JOHNES_TSIONAS:2019}, they study the well-worn Maimonides rule on class size: class sizes should be no larger than 22 students per teacher. This rule randomly creates smaller classes. Imagine a school with 22 students -- and hence one class. After a single student enters, the class is split into two, resulting in class sizes of 11 and 12. So the idea is that this threshold results in randomization of class size. By comparing schools on either side of the threshold that are otherwise identical, we can identify the impact of class size on student achievement. 

There are two main RDDs: sharp (SRD) and fuzzy (FRD). In an SRD, treatment assignment $D=1$ is a deterministic function of a single covariate (the forcing variable):
\begin{equation}
D=1\{Z\geq c\} \label{eq:force}
\end{equation}
where all participants with a value of $Z$ of at least $c$ are assigned to the treatment group -- i.e. treatment is mandatory. Alternatively, participants with $Z<c$ are ineligible for treatment. Clearly the nature of the treatment can be changed if being below some threshold results in treatment. In the \citeasnoun{JOHNES_TSIONAS:2019} example with classroom size, $c=22$ and when a school has classroom size that exceeds $c$, it is `treated' -- an additional class must be created to produce smaller classes. 

In the SRD, what is commonly identified is the average treatment effect \textbf{at} the discontinuity point (given covariates):
\begin{align}
\Delta^{ATE}_{SRD}=&E[Y(1)-Y(0)|Z=c,X=x]\notag\\=&\underset{z\downarrow c}{\lim}E[Y|Z=z,X=x]-\underset{z\uparrow c}{\lim}E[Y|Z=z,X=x]. \label{eq:SRD}
\end{align}
We are simply looking at differences in the average outcome as we approach the discontinuity from above (the right) and below (the left). 

In the SRD we impose a standard unconfoundedness assumption:
\begin{equation}
Y(0),Y(1)\perp D|Z.
\end{equation}
For identification to hold, we must further assume that overlap exists in assignment to treatment and control on both covariates and the forcing variable,
\begin{equation}
0<Pr(D=1|Z=z, X=x)<1.
\end{equation}
In this setting, this condition is fundamentally violated. For an SRD, $Pr(D=1|Z=z, X=x)$ is either 0 or 1, never in between. That is, we do not have overlap in $Z$. To identify the treatment effect extrapolation is needed. To put this another way, for $Y(0)$ ($Y(1)$) we never observe values to the right of $c$ (left of $c$). So, to identify the treatment effect it is common to make two simplifying assumptions.
\begin{assumption}
Continuity of Conditional Means of Potential Outcomes:
\begin{equation}
E[Y(0)|X=x,Z=z]\quad \text{and }\quad E[Y(1)|X=x,Z=z]
\end{equation}
are continuous in $z$. \label{a:continuity}
\end{assumption}
and
\begin{assumption}
Continuity of Conditional Distribution Functions:
\begin{equation}
F_{Y(0)|X,Z}(y|x,z)\quad \text{and }\quad F_{Y(1)|X,Z}(y|x,z)
\end{equation}
are continuous in $z$ for all $y$.
\end{assumption}
These assumptions do not place any tight restrictions on the underlying model and can be viewed as innocuous with respect to the theory.

Under these assumptions we have
\begin{align}
E[Y(0)|X=x, Z=c]=&\underset{z\uparrow c}{\lim}E[Y(0)|X=x,Z=z]=
\underset{z\uparrow c}{\lim}E[Y(0)|D=0,X=x, Z= z]\nonumber
\\=&\underset{ z\uparrow c}{\lim}E[Y|X=x, Z= z],
\end{align}
where the first equality follows from Assumption \ref{a:continuity}, the second from unconfoundedness, and the last from the fact that $Y=Y(0)$ when $D=0$. Similarly we have $E[Y(1)|X=x, Z=c]=\underset{z\downarrow c}{\lim}E[Y|X=x, Z =z]$. Accordingly, our average treatment effect when $Z=c$ is
\begin{equation}
\Delta^{ATE}_{SRD}=\underset{z\downarrow c}{\lim}E[Y|X=x,Z=z]-
\underset{z\uparrow c}{\lim}E[Y|X=x,Z=z].
\end{equation}
Our estimate of the treatment effect is the difference of two regression functions at a point. 

Now, consider the extension to the stochastic frontier model. In the setting, where treatment can effect both the production frontier and inefficiency, a simple regression setting will confound these two effects. \citeasnoun{JOHNES_TSIONAS:2019} took  a SRD approach, embedded it in a stochastic frontier framework, and obtained exactly this result in their class size investigation, noting that ``Our finding that class size influences performance in the expected direction while there is a countervailing impact on efficiency may contribute an explanation for the ambiguous results obtained in earlier research.'' Thus, as with the DiD setting, it can be misleading to interpret an estimated treatment effect directly when treatment may impact both the frontier and inefficiency. 

The difference with an SRD and an FRD is that an FRD does not make treatment assignment mandatory at the threshold. Rather, there is a non-unitary jump in the probability of assignment at the threshold: 
\begin{equation}
0\leq\underset{z\downarrow c}{\lim}Pr(D = 1|Z=z)\neq
\underset{z\uparrow c}{\lim}Pr(D = 1|Z=z)\leq 1, \label{eq:FRD}
\end{equation}
instead of $D=1\{Z\geq c\}$ that an SRD would impose. The non-unit jump in assignment probability arises if the incentives to participate in the program change in a discontinuous fashion at the threshold. In our running example, suppose that farms below a certain size threshold are eligible to apply for support to transition to organic production. Crossing the threshold qualifies a farm for application, but it does not guarantee that the farm will actually adopt organic methods -- hence, while $Pr(D=1|Z=z)=0$ when $Z<c$, once $Z>c$, $0<Pr(D=1|Z=z)<1$.

In the FRD, determination of the causal effect of treatment is not as simple as comparing regression outcomes from the left and right of the threshold. Since treatment assignment is no longer mandatory, we must account for both differences in treatment assignment and assignment around the threshold. Formally, our average treatment effect in the FRD is:
\begin{align}
\Delta^{ATE}_{FRD}=&\frac{\underset{z\downarrow c}{\lim}E[Y|X=x,Z=z]-\underset{z\uparrow c}{\lim} E[Y|X=x,Z=z]}{\underset{z\downarrow c}{\lim}E[D|X=x,Z=z]-\underset{z\uparrow c}{\lim}E[D|X=x,Z=z]}.
\end{align}

Interpreting $\Delta^{ATE}_{FRD}$ requires care. To aid in our discussion, we introduce the notation $D(z|X=x)$ to define potential treatment status given the cutoff point $z$. $D(z|X_i=x)=1$ if the $i$th farm would join the treatment group if the threshold was somehow moved from $c$ to $z$. Next, we define a complier as a farm that has 
\begin{equation}
\underset{z\downarrow Z_i}{\lim}D(z|X_i=x)=0 \quad\text{and }
\quad \underset{z\uparrow Z_i}{\lim} D(z|X_i=x)=1.
\label{eq:complier}
\end{equation}
This setting shows that assuming monotonicity is useful.
\begin{assumption}
Monotonicity of Treatment Status: $D(z|X_i=x)$ is nonincreasing in $z$ at $z=c$.
\end{assumption}
We see from Equation \eqref{eq:complier} that compliers are those farms that would take assignment to treatment if the threshold was lower than $Z_i$, but would not if the cutoff was higher than $Z_i$. That is, compliers are rule-followers. They correspond to exactly those farms who follow the intent of treatment assignment based on the level of the threshold. 

Aside from compliers we also have always-takers and never-takers. We define always-takers as
\begin{equation}
\underset{z\downarrow Z_i}{\lim}D(z|X_i=x)=1 \quad\text{and }
\quad \underset{z\uparrow Z_i}{\lim} D(z|X_i=x)=1
\label{eq:always}
\end{equation}
and never-takers as
\begin{equation}
\underset{z\downarrow Z_i}{\lim}D(z|X_i=x)=0 \quad\text{and }
\quad \underset{z\uparrow Z_i}{\lim} D(z|X_i=x)=0.
\label{eq:never}
\end{equation}
The always- and never-takers are those farms who pay no mind to the threshold. Always-takers find a way to become enrolled in the treatment regardless of their position to the threshold and never-takers essentially avoid treatment even if they qualify. In the FRD this behavior is important since always-takers and compliers show up \textbf{together} in the treatment group and compliers and never-takers show up \textbf{together} in the control. To recover an average treatment effect we need to separate these two groups on each side of the threshold.\footnote{Note there is a fourth group, defiers, but we exclude their existence since they are unlikely to exist or to exist in large enough numbers in realistic applications.}

With the three groups defined we can now specify the average treatment effect:
\begin{align}
\Delta^{ATE}_{FRD}=&\frac{\underset{z\downarrow c}{\lim}E[Y_i|X_i=x,Z_i=z]-\underset{z\uparrow c}{\lim} E[Y_i|X_i=x,Z_i=z]}{\underset{z\downarrow c}{\lim} E[D|X_i=x,Z_i=z]- \underset{z\uparrow c}{\lim}E[D|X_i=x,Z_i=z]}\nonumber\\=& E[Y_i(1)-Y_i(0)|\text{farm }i\text{ is a complier,
}X_i=x,Z_i=c].
\end{align}
This treatment effect is an average treatment effect, but it is not for all farms, only those farms that are compliers and reside \textbf{at} the threshold. Practically, there will not exist enough farms at the boundary to estimate the treatment effect with any precision. Thus, one must create a boundary which approximates this boundary residence. That is, rather than look at a farm that lies just to the left or right of the boundary, we use farms that are within some \textit{distance} of the threshold.

The conditional expectation of the observed outcome, is
\begin{align*}
E[Y|X=x,Z=z]=&E[Y(0)|D=0,X=x,Z=z]Pr(D=0|X=x,Z=z)\nonumber\\&+E[
Y(1)|D=1, X=x,Z=z]Pr(D=1|X=x,Z=z).
\end{align*}
Note that this conditional expectation lies everywhere between the individual conditional expectations; the FRD is a weighted average of two different conditional expectations, one for always-takers and another for never-takers. Identification of the policy is achieved due to the presence of the compliers that results in the discontinuity of the weighted average. That is, the policy is evaluated for compliers near the boundary/threshold.

While the SRD is a weighted average as well, the intuition behind it differs from the FRD. In the SRD, treatment assignment is \textbf{deterministic} at the threshold: crossing the cutoff induces a discrete change in treatment status for all units. In this sense, all units near the threshold can be treated as effectively compliers in the sense that treatment status is fully determined by the cutoff.

A simple and widely used way to implement an SRD design is through a local linear regression in a neighborhood of the threshold. Restricting attention to observations sufficiently close to the cutoff, one may write
\begin{equation}
Y_i = \alpha + D_i\beta_1  +  (Z_i - c)\beta_2 +  D_i (Z_i - c) \beta_3 + \varepsilon_i,
\end{equation}
where $D_i = 1\{Z_i \geq c\}$ indicates treatment status, $Z_i$ is the running variable, and $c$ is the cutoff. The parameter $\beta_1$ captures the discontinuity in the outcome at the threshold and therefore represents the RDD treatment effect, while $\beta_3$ allows the slope in the running variable to differ on either side of the cutoff.

In practice, estimation requires restricting the sample to observations sufficiently close to the threshold, with the notion of ``closeness'' governed by a bandwidth choice. Because this choice is inherently empirical, it is important to assess the robustness of estimated treatment effects to alternative bandwidths.

The same structure carries over to the stochastic frontier setting. Mimicking the equal-timing DiD of Section~4, the model is
\begin{align*}
Y_{i}=&\alpha+D_i\beta_1+(Z_i-c)\beta_2+D_i(Z_i-c)\beta_{3}+ X_i\gamma+v_i-u_i\\
=&\alpha+D_i\beta_1+(Z_i-c)\beta_2+D_i(Z_i-c)\beta_{3}+X_i\gamma+\varepsilon_i,
\end{align*}
which looks identical to the pure regression setup except that now we have explicitly included inefficiency in the model and have added in traditional inputs of production $X$. However, at this point the researcher needs to make a decision if the threshold effect \textbf{only} materializes through the frontier technology (direct) or if it also impacts inefficiency (indirect). This can be modeled by including the same three covariates in the mean of the distribution of $u_i$. 

One must exert care here because a key assumption that has been overlooked in the SFA-RDD literature to date \cite{JOHNES_TSIONAS:2019,BADUNENKO_ETAL:2023} is if the threshold impacts the full distribution of inefficiency or only a part of it. What we mean is the following. The original SFA-RDD paper by \citeasnoun{JOHNES_TSIONAS:2019} assumed that $u\sim N_+(\mu,\sigma_u^2)$ and $\mu$ was modeled as $\mu=\rho_0+D_i\rho_1+(Z_i-c)\rho_2+D_i(Z_i-c)\rho_3$ but the variance, $\sigma_u^2$ was kept constant. This assumption does allow the distribution of $u$ to change at the threshold, but it is restrictive because in truncated distributions the moments usually depend on all of the parameters. Thus, the mean of $u$ in the \citeasnoun{JOHNES_TSIONAS:2019} framework could have also been shifted at the threshold by specifying $\ln \sigma^2_u=\rho_0+D_i\rho_1+(Z_i-c)\rho_2+D_i(Z_i-c)\rho_3$, or both parameters could have had threshold changes modeled in them. Moreover, it is also the case that the distribution of inefficiency could change as a farm crosses the threshold and enters treatment. 

One simple way that the parameter specification issue could be avoided in the SFA-RDD setting would be to assume that inefficiency comes from a distribution that satisfies the scaling property \cite{WANG_SCHMIDT:2002}. In this case we may write the distribution of $u$ as $g(S,\delta)u^*_i$ where $g(S,\delta)\ge 0$, $S$ is an observable random vector and $u^*$ has a distribution that does not depend on $Z$ (and by consequence is invariant to the position of the farm and the threshold). $g(S,\delta)$ is commonly called the scaling function. The only real requirement on it is that it is nonnegative. All one parameter families of distributions (Half-Normal, Exponential, Generalized Exponential) automatically possess the scaling property (and this is what \citeasnoun{BADUNENKO_ETAL:2023} assume), but many other distributions can be made to possess this property. For example, using the \citeasnoun{JOHNES_TSIONAS:2019} setting, if $u\sim N_+(\mu,\sigma_u^2)$, we could enforce the scaling property as $u \vert S = s\sim N_+(g(s,\delta)\mu,g(s,\delta)^2\sigma_u^2)$. 

A benefit of invoking/assuming the scaling property is that now the assumption is that the scaling function is what is impacted by the crossing of the threshold (treatment assignment) but not the basic distribution of inefficiency. Another appealing feature is that the RDD model could be couched as a nonlinear regression model. Setting $g(S_i,\rho)=exp\left(\rho_0+D_i\rho_1+(Z_i-c)\rho_2+D_i(Z_i-c)\rho_3\right)$, our SFA-RDD model is now
\begin{align}
Y_{i}=&\alpha+D_i\beta_1+(Z_i-c)\beta_2+D_i(Z_i-c)\beta_{3}+ X_i\gamma+v_i-u_i\notag\\
=&\alpha+D_i\beta_1+(Z_i-c)\beta_2+D_i(Z_i-c)\beta_{3}+ X_i\gamma-E[u_i]+v_i-\left(u_i-E[u_i]\right)\notag\\
=&\alpha+D_i\beta_1+(Z_i-c)\beta_2+D_i(Z_i-c)\beta_{3}+ X_i\gamma-\mu^*g(S_i,\rho)+\varepsilon^*_i.
\end{align}
where $\mu^*=E[u^*]$ is the mean of the basic distribution and $\varepsilon^*_i$ is now a 0 mean random variable. This model is a nonlinear version of standard RDD. 

If one did not wish to make such an assumption and follow the empirical work of \citeasnoun{JOHNES_TSIONAS:2019}, then a distribution would need to be specified for $u$ (Truncated Normal for example) and then the parameter(s) would need to be specified as a function of $S=\left(D_i,Z_i-c,D_i(Z_i-c)\right)$. 

The key empirical issue is how to select those observations that are close to the threshold that make up the treatment and control groups. A standard approach is to run a nonparametric regression of the output on the running variable \cite{IMBENS_KALYANARAMAN:2012}. The bandwidth that is used for this regression controls the bias-variance trade off. In essence, a local linear regression is run on each side of the threshold. On each side of the threshold, for the running variable $Z_i$, we estimate a local linear regression 
\begin{align*}
E[Y_i|Z_i=c]\approx&\alpha_0+(Z_i-c)\beta_0\quad \text{for }Z_i<c\\
E[Y_i|Z_i=c]\approx&\alpha_1+(Z_i-c)\beta_1\quad \text{for }Z_i\ge c
\end{align*}
for $Z_i\in\left[c-h,c+h\right]$. \citeasnoun{IMBENS_KALYANARAMAN:2012} suggest choosing the bandwidth $h$ such that the MSE is minimized, where $MSE=Bias^2+Var$. Here the $MSE$ is defined over the treatment effect. What is interesting is that the bias and variance must be estimated in order to select $h$ to minimize MSE. \citeasnoun{IMBENS_KALYANARAMAN:2012} suggest using the full sample and a pilot bandwidth to estimate the corresponding terms. From there a closed form solution exists for the optimal $h$, which can then be used. 

An improvement to \possessivecite{IMBENS_KALYANARAMAN:2012} approach is that of \citeasnoun{CALONICO_ETAL:2014}, who improve on the bias features of their procedure. The approach of \citeasnoun{CALONICO_ETAL:2014} essentially bias corrects the estimate of the treatment effect and allows for valid inference. Thus, both approaches allow you to select the window near the threshold to estimate the treatment effect, but \citeasnoun{CALONICO_ETAL:2014} corrects the bias inherent in this estimator and also provides valid confidence intervals. 

Notice the difference between the regression discontinuity regression and the DiD regression. In the DiD regression the change in one group's status over time is exploited to identify the treatment effect. In the RDD, a discrete policy change, coupled with continuous varying spatial effects allows one to pin down the impact of the treatment. In essence, DiD exploits time-space variation while RDD exploits a discrete jump over spatial variation.

See \citeasnoun{IMBENS_LEMIEUX:2008} as well as the corresponding special
issue of the \textit{Journal of Econometrics} for more on both
theory and applications of regression discontinuity. It is important to point out that mandatory treatment status may not be optimal. In the FRD compliers are those that \textit{choose} to receive treatment --  they have some belief that treatment will be beneficial to them. A policy that mandates assignment may include farms that do not want to receive treatment and prevent those who want it from getting it. Here we do not discuss optimality issues with treatment assignment other than to say that when one has compliers, selection effects need to be controlled for to adequately estimate the treatment effect.

\section{Outstanding Open Problems in Causal SFA}

While the preceding sections demonstrate that many modern causal inference tools can be readily adapted to the stochastic frontier setting, the literature remains in its infancy. Indeed, many of the approaches surveyed above represent only first steps toward integrating quasi-experimental design with the composed-error structure of stochastic frontier models. Several important conceptual, econometric, and computational questions remain unresolved. We briefly highlight what we view as some of the most promising directions for future research.

\subsection{Parallel Trends and Diagnostic Testing in
SFA Difference-in-Differences}\label{sec:open-parallel}

A particularly important unresolved issue concerns the formulation and empirical validation of parallel trends assumptions in stochastic frontier difference-in-differences settings. In standard DiD applications, the parallel trends assumption pertains to the untreated evolution of observed outcomes. In the SFA setting, however, treatment may separately affect both the production frontier \textbf{and} the inefficiency distribution, raising the question of whether identification requires parallel trends in each component individually or only in their composite. Put differently, it remains unclear what precise restrictions must hold on the latent frontier and inefficiency processes in order for causal decomposition to be valid.

Relatedly, because inefficiency is latent, standard pre-treatment diagnostic tools for assessing parallel trends may not directly apply. Event-study plots and placebo regressions may reveal whether observed output evolves similarly prior to treatment, but they do not separately diagnose whether pre-treatment trends differ in frontier shifts versus inefficiency movements. Developing formal identification results, specification tests, and robustness diagnostics for parallel trends in composed-error environments remains an important open methodological challenge.

\subsection{Synthetic Control Methods in Stochastic Frontier Settings}

Another promising direction concerns the extension of synthetic control methods to stochastic frontier models. As discussed previously, while synthetic control methods can in principle recover reduced-form treatment effects on observed output, adapting them to settings where treatment may affect both the frontier and inefficiency distribution remains difficult. Existing synthetic control approaches are designed to match observed outcomes rather than latent decompositions of the outcome-generating process.

Future research may seek to develop synthetic control procedures that incorporate estimated frontier and inefficiency components directly into the matching criterion, or alternatively construct synthetic controls within the likelihood function itself. Such approaches could prove especially valuable in applications with a small number of treated units and long pre-treatment panels, where standard DiD methods may be less appropriate.

\subsection{Semiparametric and Nonparametric Causal SFA}

Nearly all existing work at the intersection of causal inference and SFA remains firmly parametric, relying on strong functional form and distributional assumptions regarding both the production frontier and inefficiency distribution. While such assumptions are standard in classical SFA, they may become increasingly restrictive when paired with quasi-experimental identification assumptions.

Accordingly, an important frontier for future work lies in developing semi- and nonparametric estimators for causal stochastic frontier models. Potential directions include local polynomial or kernel-based DiD estimators for frontier models, semiparametric regression discontinuity estimators with latent inefficiency, and partially linear formulations that relax the parametric structure of either the frontier or treatment-response function (see \citeasnoun{SIMAR_VANKEILEGOM_ZELENYUK:2017}, \citeasnoun{FLORENS_SIMAR_VANKEILEGOM:2020},  \citeasnoun{CENTORRINO_PARMETER:2024}, \citeasnoun{NieParmeterZelenyukZhang2025}, \citeasnoun{WangSunKumbhakar2025} among others). Such developments would align naturally with the broader movement toward semiparametric identification in modern econometrics.

\subsection{Multi-Output and Distance Function Extensions}

To date, most causal SFA work has focused on single-output production frontier models. Yet many empirical applications in productivity and efficiency analysis involve multiple outputs, multiple inputs, or dual representations of technology such as cost, revenue, and profit functions \cite{ASSAF_ATKINSON_TSIONAS:2020,DINVERNO_SMET_DEWITTE:2021,TSIONAS_IZZELDIN_HENNINGSEN_PARAVALOS:2022}. Extending causal inference tools to these environments remains largely unexplored.

In particular, integrating DiD or RDDs into directional distance functions, stochastic ray production frontiers, or cost frontier models would substantially broaden the applicability of the methods discussed in this chapter. These settings may also provide additional identifying information by exploiting duality relationships and alternative economic restrictions. Developing causal decomposition results in such frameworks therefore remains a promising avenue for future research.

\subsection{Dynamic and Continuous Treatment Frameworks}

The causal SFA literature has thus far focused almost exclusively on binary treatment settings, where farms or producers are either treated or untreated. In practice, however, many policy interventions are dynamic, repeated, or continuous in nature. Examples include phased regulatory reforms, gradually increasing subsidy levels, and treatments defined by intensity rather than binary participation.

Future research should therefore consider extending causal SFA methods to accommodate dynamic treatment effects, continuous treatment regimes, and dose-response relationships. This may include adapting generalized propensity score methods, dynamic DiD estimators, or distributed lag frameworks to stochastic frontier settings. Such extensions would significantly expand the empirical relevance of causal SFA beyond the canonical binary-treatment case.

\subsection{Machine Learning and High-Dimensional Covariate Adjustment}

Recent years have witnessed growing interest in incorporating machine learning tools into stochastic frontier analysis. Existing contributions include regularization-based approaches for variable selection and high-dimensional frontier estimation \cite{jin/lee:18,horrace/etal:23,PARMETER_ETAL:2025}, as well as neural-network and deep-learning methods \cite{KUTLU_ETAL:2026,WEI_ETAL:2026} for flexibly approximating production technologies. These developments suggest that the frontier literature is beginning to engage seriously with modern predictive methodologies.

Nevertheless, the integration of machine learning methods into explicitly causal stochastic frontier settings remains largely unexplored. In many empirical applications, treatment assignment mechanisms and nuisance components may be high-dimensional or highly nonlinear, creating scope for modern debiased and regularized estimation methods. Potential directions for future work include adapting double machine learning procedures for treatment estimation in frontier models, employing regularization methods for covariate selection in causal SFA specifications, and embedding flexible machine learning approximators within quasi-experimental frontier frameworks. Extending existing machine learning frontier methods to settings involving treatment effects and causal decomposition remains a promising direction for future research.

\section{Conclusions}

The importance of causal inference methods in applied economics is now beyond debate. The ability to identify quasi-experimental variation has become central to uncovering credible causal effects across a wide range of applied milieux. Yet one area in which these methods have only recently begun to permeate is the production and efficiency literature. This omission is notable, as many policies and institutional changes may plausibly affect both the production technology itself and the productive efficiency of farms. While some recent work, particularly in the education production literature, has begun to explore this intersection, a comprehensive treatment has until now remained absent. We hope this chapter provides a useful foundation for future work at the intersection of causal inference and SFA.


In general, combining causal inference methods and frontier methods represents a new frontier (pun not intended) and one that has the ability to generate many new insights regarding regulatory policies that may/may not harm the ability of producers to improve their performance.

A unifying theme of this chapter is that the composed-error structure of the stochastic frontier model -- $\varepsilon_i = v_i - u_i$, where $v_i$ is symmetric noise and $u_i \geq 0$ is inefficiency -- systematically complicates the direct application of causal inference methods from standard regression. A causal intervention may affect \emph{both} the production frontier and the inefficiency distribution, so that reduced-form treatment effects generally conflate direct and indirect channels that are of independent economic interest.

Researchers seeking to combine causal inference methods with SFA must therefore remain mindful of these challenges. As we have argued throughout, many of these difficulties can be addressed through careful econometric design and the use of distributional assumptions that are already familiar within the broader SFA literature.


\bibliographystyle{agsm} 
\bibliography{modeval}

@article{TRAN2015,
	author = {Kien C. Tran and Efthymios G. Tsionas},
	doi = {https://doi.org/10.1016/j.econlet.2015.05.026},
	issn = {0165-1765},
	journal = {Economics Letters},
	pages = {85-88},
	title = {Endogeneity in stochastic frontier models: Copula approach without external instruments},
	url = {https://www.sciencedirect.com/science/article/pii/S0165176515002219},
	volume = {133},
	year = {2015}}

@article{jin/lee:18,
	article-number = {8},
	author = {Jin, Fei and Lee, Lung-fei},
	doi = {10.3390/econometrics6010008},
	issn = {2225-1146},
	journal = {Econometrics},
	number = {1},
	title = {Lasso Maximum Likelihood Estimation of Parametric Models with Singular Information Matrices},
	url = {https://www.mdpi.com/2225-1146/6/1/8},
	volume = {6},
	year = {2018}}

@article{horrace/etal:23,
	author = {William C. Horrace and Hyunseok Jung and Yoonseok Lee},
	doi = {10.1080/07350015.2022.2110881},
	journal = {Journal of Business \& Economic Statistics},
	number = {0},
	pages = {1-11},
	publisher = {Taylor & Francis},
	title = {LASSO for Stochastic Frontier Models with Many Efficient Firms},
	volume = {0},
	year = {2022}}

@article{WEI_ETAL:2026,
	author = {Wei, Zheng and Sang, Huiyan and Prokhorov, Artem and Ma, Yu},
	doi = {10.1007/s11123-026-00797-3},
	journal = {Journal of Productivity Analysis},
	number = {2},
	pages = {1--16},
	title = {Shape-aware deep learning for models of production},
	volume = {65},
	year = {2026}}

@article{KUTLU_ETAL:2026,
	author = {Kutlu, Levent and Mao, Xi and Ni, Sherry},
	journal = {Journal of Productivity Analysis},
	note = {forthcoming},
	title = {A Machine Learning Approach to Stochastic Frontier Modeling},
	year = {2026}}

@article{PARMETER_ETAL:2025,
	author = {Christopher F. Parmeter and Artem Prokhorov and Valentin Zelenyuk},
	journal = {arXiv preprint arXiv:2505.14282},
	title = {The Post Double {LASSO} for Efficiency Analysis},
	year = {2025}}

@book{AZZALINI:2014,
	author = {Azzalini, A.},
	publisher = {Cambridge University Press},
	title = {The Skew-normal and Related Families},
	year = {2014}}

@book{CUNNINGHAM:2021,
	address = {London},
	author = {Scott Cunningham},
	publisher = {Yale University Press},
	title = {Causal Inference: The Mixtape},
	year = {2021}}

@book{KLEIN:2021,
	author = {Huntington-Klein, N.},
	publisher = {Chapman and Hall/CRC},
	title = {The Effect: An Introduction to Research Design and Causality},
	year = {2021}}

@article{BADUNENKO_ETAL:2023,
	author = {Oleg Badunenko and Giovanna D'Inverno and Kristof {De Witte}},
	journal = {European Journal of Operational Research},
	number = {1},
	pages = {432-447},
	title = {On distinguishing the direct causal effect of an intervention from its efficiency-enhancing effects},
	volume = {310},
	year = {2023}}

@article{CALLAWAY_LI:2023,
	author = {Brantly Callaway and Tong Li},
	doi = {https://doi.org/10.1016/j.jeconom.2023.03.009},
	issn = {0304-4076},
	journal = {Journal of Econometrics},
	number = {1},
	pages = {105454},
	title = {Policy evaluation during a pandemic},
	url = {https://www.sciencedirect.com/science/article/pii/S0304407623001483},
	volume = {236},
	year = {2023}}

@article{GOODMAN-BACON_MARCUS:2020,
	author = {Goodman-Bacon, Andrew and Marcus, Jan},
	doi = {10.18148/srm/2020.v14i2.7723},
	journal = {Survey Research Methods},
	month = {Jun},
	number = {2},
	title = {Using Difference-in-Differences to Identify Causal Effects of COVID-19 Policies},
	url = {https://ojs.ub.uni-konstanz.de/srm/article/view/7723},
	volume = {14},
	year = {2020}}

@article{HENDERSON:2022,
	author = {Heath Henderson},
	journal = {Journal of Human Development and Capabilities},
	number = {3},
	pages = {425--454},
	title = {The Moral Foundations of Impact Evaluation},
	volume = {23},
	year = {2022}}

@article{MILLIMET_PARMETER:2022,
	author = {Millimet, Daniel L. and Parmeter, Christopher F.},
	journal = {Political Analysis},
	number = {1},
	title = {Accounting for Skewed or One-Sided Measurement Error in the Dependent Variable},
	volume = {30},
	year = {2022}}

@article{IMBENS_LEMIEUX:2008,
	author = {Guido W. Imbens and Thomas Lemieux},
	journal = {Journal of Econometrics},
	number = {2},
	pages = {615-635},
	title = {Regression discontinuity designs: A guide to practice},
	volume = {142},
	year = {2008}}

@article{CENTORRINO_PEREZ_URETA_WALL:2024,
	author = {Centorrino, Samuele and P{\'e}rez-Urdiales, Mar{\'\i}a and Bravo-Ureta, Boris and Wall, Alan},
	journal = {Journal of Applied Econometrics},
	number = {3},
	pages = {365-382},
	title = {Binary endogenous treatment in stochastic frontier models with an application to soil conservation in {El Salvador}},
	volume = {39},
	year = {2024}}

@article{CENTORRINO_PEREZ:2019,
	author = {Samuele Centorrino and Mar{\'\i}a P{\'e}rez-Urdiales},
	issn = {0304-4076},
	journal = {Journal of Econometrics},
	number = {1},
	pages = {82-105},
	title = {Maximum likelihood estimation of stochastic frontier models with endogeneity},
	volume = {234},
	year = {2023}}

@unpublished{TO-THEA_TUAN:2019,
	author = {{To-The}, Nguyen and Tuan, {Nguyen Anh}},
	note = {SSRN working paper 3361741},
	title = {Benchmarking Technology in the Stochastic Frontier Maize Production of Northern {Vietnam}: Difference in Difference Approach},
	url = {https://ssrn.com/abstract=3361741},
	year = {2019}}

@article{IMBENS_KALYANARAMAN:2012,
	author = {Imbens, Guido W and Kalyanaraman, Karthik},
	doi = {10.1093/restud/rdr043},
	journal = {The Review of Economic Studies},
	number = {3},
	pages = {933--959},
	publisher = {Oxford University Press},
	title = {Optimal Bandwidth Choice for the Regression Discontinuity Estimator},
	url = {https://doi.org/10.1093/restud/rdr043},
	volume = {79},
	year = {2012}}

@article{CALONICO_ETAL:2014,
	author = {Calonico, Sebastian and Cattaneo, Matias D and Titiunik, Rocio},
	doi = {10.3982/ECTA11757},
	journal = {Econometrica},
	number = {6},
	pages = {2295--2326},
	publisher = {Wiley},
	title = {Robust nonparametric confidence intervals for regression?discontinuity designs},
	url = {https://doi.org/10.3982/ECTA11757},
	volume = {82},
	year = {2014}}

@article{JOHNES_TSIONAS:2019,
	author = {Johnes, Geraint and Tsionas, Mike G.},
	doi = {https://doi.org/10.1002/sta4.242},
	eprint = {https://onlinelibrary.wiley.com/doi/pdf/10.1002/sta4.242},
	journal = {Stat},
	number = {1},
	pages = {e242},
	title = {A regression discontinuity stochastic frontier model with an application to educational attainment},
	url = {https://onlinelibrary.wiley.com/doi/abs/10.1002/sta4.242},
	volume = {8},
	year = {2019}}

@article{CALLAWAY_SANTANA:2021,
	author = {Callaway, Brantly and Sant'Anna, Pedro HC},
	journal = {Journal of Econometrics},
	number = {2},
	pages = {200--230},
	publisher = {Elsevier},
	title = {Difference-in-differences with multiple time periods},
	volume = {225},
	year = {2021}}

@article{DINVERNO_ETAL:2023,
	author = {Giovanna D'Inverno and Francesco Vidoli and Kristof {De Witte}},
	doi = {https://doi.org/10.1016/j.ejor.2023.01.036},
	issn = {0377-2217},
	jourl = {European Journal of Operational Research},
	number = {2},
	pages = {857-871},
	title = {Sustainable budgeting and financial balance: Which lever will you pull?},
	url = {https://www.sciencedirect.com/science/article/pii/S037722172300067X},
	volume = {309},
	year = {2023}}

@book{ANGRIST_PISCHKE:2014,
	author = {Joshua D. Angrist and Jorn-Steffen Pischke},
	edition = {1},
	publisher = {Princeton University Press},
	series = {Economics Books},
	title = {Mastering 'Metrics: The Path from Cause to Effect},
	year = {2014}}

@article{SUN_ABRAHAM:2021,
	author = {Liyang Sun and Sarah Abraham},
	doi = {https://doi.org/10.1016/j.jeconom.2020.09.006},
	issn = {0304-4076},
	journal = {Journal of Econometrics},
	note = {Themed Issue: Treatment Effect 1},
	number = {2},
	pages = {175-199},
	title = {Estimating dynamic treatment effects in event studies with heterogeneous treatment effects},
	url = {https://www.sciencedirect.com/science/article/pii/S030440762030378X},
	volume = {225},
	year = {2021}}

@article{KUTLU:2010,
	author = {Kutlu, L.},
	journal = {Economics Letters},
	pages = {79-81},
	title = {{Battese-Coelli} estimator with endogenous regressors},
	volume = {109},
	year = {2010}}

@article{KARAKAPLAN_KUTLU:2017,
	author = {Karakaplan, M. U. and Kutlu, L.},
	journal = {Economics Bulletin},
	pages = {889-901},
	title = {Handling endogeneity in stochastic frontier analysis},
	volume = {37},
	year = {2017}}

@article{KUTLU_ETAL:2020,
	author = {Levent Kutlu and Kien C. Tran and Mike G. Tsionas},
	journal = {European Journal of Operational Research},
	number = {1},
	pages = {389-399},
	title = {A spatial stochastic frontier model with endogenous frontier and environmental variables},
	volume = {286},
	year = {2020}}

@article{TRAN_TSIONAS:2013,
	author = {Tran, K. C. and Tsionas, E. G.},
	journal = {Economics Letters},
	pages = {233-236},
	title = {{GMM} estimation of stochastic frontier models with endogenous regressors},
	volume = {118},
	year = {2013}}

@article{AMSLER_PROKHOROV_SCHMIDT:2016,
	author = {Amsler, C. and Prokhorov, A. and Schmidt, P.},
	journal = {Journal of Econometrics},
	pages = {280-288},
	title = {Endogeneity in stochastic frontier models},
	volume = {190},
	year = {2016}}

@article{AMSLER_PROKHOROV_SCHMIDT:2017,
	author = {Amsler, C. and Prokhorov, A. and Schmidt, P.},
	journal = {Journal of Econometrics},
	pages = {131-140},
	title = {Endogenous environmental variables in stochastic frontier models},
	volume = {199},
	year = {2017}}

@article{KARANKI_LIM:2024,
	author = {Karanki, F. and Lim, S. H.},
	journal = {Transportation Research Record},
	number = {12},
	pages = {673-686},
	title = {Role of {U.S.} Airports' Attributes in the Output Response to the {COVID-19} Pandemic: A Stochastic Frontier Analysis with the Difference-in-Differences Method},
	volume = {2678},
	year = {2024}}

@incollection{PARMETER:2023,
	address = {Cham},
	author = {Parmeter, C. F.},
	booktitle = {Advanced Mathematical Methods for Economic Efficiency Analysis: Theory and Empirical Applications},
	editor = {Macedo, Pedro and Moutinho, Victor and Madaleno, Mara},
	pages = {229--249},
	publisher = {Springer International Publishing},
	title = {Is it {MOLS} or {COLS}?},
	year = {2023}}

@article{FLORENS_SIMAR_VANKEILEGOM:2020,
	author = {Florens, J.-P. and Simar, L. and {Van} Keilegom, I.},
	journal = {Journal of the American Statistical Association},
	pages = {425-441},
	title = {Estimation of the Boundary of a Variable Observed With Symmetric Error},
	volume = {115},
	year = {2020}}

@article{CENTORRINO_PARMETER:2024,
	author = {Samuele Centorrino and Christopher F. Parmeter},
	journal = {Journal of Econometrics},
	number = {2},
	pages = {105641},
	title = {Nonparametric estimation of stochastic frontier models with weak separability},
	volume = {238},
	year = {2024}}

@article{CHEN_ETAL:2014,
	author = {Chen, Y.-Y. and Schmidt, P. and Wang, H.-J.},
	journal = {Journal of Econometrics},
	number = {1},
	pages = {65-76},
	title = {Consistent estimation of the fixed effects stochastic frontier model},
	volume = {181},
	year = {2014}}

@article{CHEN_ETAL:2020,
	author = {Chen, Y.-T. and Hsu, Y.-C. and Wang, H.-J.},
	journal = {Journal of Business \& Economic Statistics},
	number = {2},
	pages = {243-256},
	title = {{A Stochastic Frontier Model with Endogenous Treatment Status and Mediator}},
	volume = {38},
	year = {2020}}

@article{BELOTTI_ILARDI:2018,
	author = {Belotti, F. and Ilardi, G.},
	journal = {Journal of Econometrics},
	number = {2},
	pages = {161-177},
	title = {Consistent inference in fixed-effects stochastic frontier models},
	volume = {202},
	year = {2018}}

@article{GREENE:2005,
	author = {Greene, W. H.},
	journal = {Journal of Econometrics},
	number = {2},
	pages = {269-303},
	title = {Reconsidering heterogeneity in panel data estimators of the stochastic frontier model},
	volume = {126},
	year = {2005}}

@article{WANG_SCHMIDT:2002,
	author = {Wang, H.-J. and Schmidt, P.},
	journal = {Journal of Productivity Analysis},
	pages = {129-144},
	title = {One-step and two-step estimation of the effects of exogenous variables on technical efficiency levels},
	volume = {18},
	year = {2002}}

@article{OLSON_SCHMIDT_WALDMAN:1980,
	author = {Olson, J. A. and Schmidt, P. and Waldman, D. A.},
	journal = {Journal of Econometrics},
	pages = {67-82},
	title = {A {Monte Carlo} study of estimators of stochastic frontier production functions},
	volume = {13},
	year = {1980}}

@article{HANSEN_DONALD_NEWEY:2010,
	author = {Hansen, C. and {McDonald}, J. B. and Newey, W. K.},
	journal = {Journal of Business and Economic Statistics},
	pages = {13-25},
	title = {Instrumental variables estimation with flexible distributions},
	volume = {28},
	year = {2010}}

@incollection{KUMBHAKAR_PARMETER_ZELENYUK:2021,
	author = {Kumbhakar, S. C. and Parmeter, C. F. and Zelenyuk, V.},
	booktitle = {Handbook of Production Economics},
	editor = {Subhash Ray and Robert Chambers and Subal C. Kumbhakar},
	publisher = {Springer},
	title = {Stochastic Frontier Analysis: Foundations and Advances {I}},
	volume = {1},
	year = {2022}}

@article{PARMETER_KUMBHAKAR:2014,
	author = {Parmeter, C. F. and Kumbhakar, S. C.},
	journal = {Foundations and Trends in Econometrics},
	number = {3-4},
	pages = {191-385},
	title = {{Efficiency Analysis: A Primer on Recent Advances}},
	volume = {7},
	year = {2014}}

@article{SIMAR_VANKEILEGOM_ZELENYUK:2017,
	author = {Simar, L. and Van{ Keilegom}, I. and Zelenyuk, V.},
	journal = {Journal of Productivity Analysis},
	number = {3},
	pages = {189-204},
	title = {Nonparametric Least Squares Methods for Stochastic Frontier Models},
	volume = {47},
	year = {2017}}

@article{Abadie2021JEL,
	author = {Abadie, Alberto},
	doi = {10.1257/jel.20191450},
	journal = {Journal of Economic Literature},
	number = {2},
	pages = {391--425},
	title = {Using Synthetic Controls: Feasibility, Data Requirements, and Methodological Aspects},
	volume = {59},
	year = {2021}}

@unpublished{BakerEtal2025,
	author = {Baker, Andrew C. and Callaway, Brantly and Cunningham, Scott and Goodman-Bacon, Andrew and Sant'Anna, Pedro H. C.},
	note = {Working paper, arXiv preprint arXiv:2503.13323},
	title = {Difference-in-Differences: A Practitioner's Guide},
	url = {https://arxiv.org/abs/2503.13323},
	year = {2025}}

@article{Greene2010,
	author = {Greene, William},
	doi = {10.1007/s11123-009-0159-1},
	journal = {Journal of Productivity Analysis},
	number = {1},
	pages = {15--24},
	title = {A Stochastic Frontier Model with Correction for Sample Selection},
	volume = {34},
	year = {2010}}

@incollection{HajargashtGriffiths2019,
	author = {Hajargasht, Gholamreza and Griffiths, William E.},
	booktitle = {Advances in Econometrics},
	publisher = {Emerald Publishing},
	title = {Estimation of Stochastic Frontier Models by a Semiparametric Approach},
	volume = {41},
	year = {2019}}

@article{Haschka2024,
	author = {Haschka, Rouven E.},
	doi = {10.1007/s10260-024-00750-4},
	journal = {Statistical Methods \& Applications},
	number = {3},
	pages = {807--826},
	title = {Endogeneity in Stochastic Frontier Models with ``Wrong'' Skewness: Copula Approach without External Instruments},
	volume = {33},
	year = {2024}}

@article{KaragiannisKellermann2019,
	author = {Karagiannis, Giannis and Kellermann, Konrad L.},
	doi = {10.1007/s11123-019-00558-z},
	journal = {Journal of Productivity Analysis},
	pages = {1--11},
	title = {Stochastic Frontier Models with Correlated Effects},
	volume = {52},
	year = {2019}}

@article{Karakaplan2022,
	author = {Karakaplan, Mustafa U.},
	doi = {10.1177/1536867X221140002},
	journal = {Stata Journal},
	number = {4},
	pages = {846--858},
	title = {Fitting Endogenous Stochastic Frontier Models in {Stata}},
	volume = {22},
	year = {2022}}

@article{NieParmeterZelenyukZhang2025,
	author = {Nie, Puguang and Parmeter, Christopher F. and Zelenyuk, Valentin and Zhang, Xibin},
	doi = {10.1016/j.jeconom.2025.106081},
	journal = {Journal of Econometrics},
	note = {Forthcoming},
	pages = {106081},
	title = {Bayesian Estimation of a Semiparametric Stochastic Frontier Model with Persistent and Transient Inefficiencies},
	year = {2025}}

@article{RothEtal2023,
	author = {Roth, Jonathan and Sant'Anna, Pedro H. C. and Bilinski, Alyssa and Poe, John},
	doi = {10.1016/j.jeconom.2023.03.008},
	journal = {Journal of Econometrics},
	number = {2},
	pages = {2218--2244},
	title = {What's Trending in Difference-in-Differences? {A} Synthesis of the Recent Econometrics Literature},
	volume = {235},
	year = {2023}}

@article{WangSunKumbhakar2025,
	author = {Wang, Taining and Sun, Kai and Kumbhakar, Subal C.},
	doi = {10.1007/s00181-024-02708-7},
	journal = {Empirical Economics},
	number = {6},
	pages = {2477--2514},
	title = {A New Semiparametric Stochastic Frontier Model: Addressing Inefficiency and Model Flexibility Using Panel Data},
	volume = {68},
	year = {2025}}

@article{BRAVO_URETA_ETAL:2020,
	author = {Bravo-Ureta, Boris E. and Gonz{\'a}lez-Flores, Mario and Greene, William and Sol{\'\i}s, Daniel},
	doi = {10.1111/ajae.12112},
	journal = {American Journal of Agricultural Economics},
	number = {1},
	pages = {362--385},
	title = {Technology and Technical Efficiency Change: Evidence from a Difference in Differences Selectivity Corrected Stochastic Production Frontier Model},
	volume = {103},
	year = {2021}}

@article{ASSAF_ATKINSON_TSIONAS:2020,
	author = {Assaf, A. George and Atkinson, Scott E. and Tsionas, Mike G.},
	doi = {10.1016/j.tourman.2020.104124},
	journal = {Tourism Management},
	pages = {104124},
	title = {Endogeneity in multiple output production: Evidence from the {US} hotel industry},
	volume = {80},
	year = {2020}}

@article{DINVERNO_SMET_DEWITTE:2021,
	author = {D'Inverno, Giovanna and Smet, Mike and De Witte, Kristof},
	doi = {10.1016/j.ejor.2020.08.042},
	journal = {European Journal of Operational Research},
	number = {3},
	pages = {1111--1124},
	title = {Impact evaluation in a multi-input multi-output setting: Evidence on the effect of additional resources for schools},
	volume = {290},
	year = {2021}}

@article{TSIONAS_IZZELDIN_HENNINGSEN_PARAVALOS:2022,
	author = {Tsionas, Mike G. and Izzeldin, Marwan and Henningsen, Arne and Paravalos, Evaggelos},
	doi = {10.1007/s00181-021-02060-0},
	journal = {Empirical Economics},
	number = {3},
	pages = {1345--1363},
	title = {Addressing endogeneity when estimating stochastic ray production frontiers: a {Bayesian} approach},
	volume = {62},
	year = {2022}}

@article{KUMBHAKAR_TSIONAS_SIPILAINEN:2009,
	author = {Kumbhakar, Subal C. and Tsionas, Efthymios G. and Sipil{\"a}inen, Timo},
	doi = {10.1007/s11123-008-0081-y},
	journal = {Journal of Productivity Analysis},
	number = {3},
	pages = {151--161},
	title = {Joint estimation of technology choice and technical efficiency: an application to organic and conventional dairy farming},
	volume = {31},
	year = {2009}}

\end{document}